\documentclass[a4paper,11pt]{article}

\usepackage{a4wide}
\usepackage{setspace}
\usepackage{graphicx}
\usepackage{xcolor}
\usepackage{cite}
\usepackage{soul}

\DeclareGraphicsExtensions{.eps,.pdf}

\renewcommand{\hl}{}

\begin{document}

\begin{center}
\Large{
Off-resonant Coherent Electron Transport over Three Nanometers in Multi-heme Protein Bioelectronic Junctions} \\[1em]
\normalsize{
Zdenek Futera$^{a,b}$, 
Ichiro Ide$^c$,
Ben Kayser$^d$
Kavita Garg$^d$
Xiuyun Jiang$^b$, 
Jessica H. van Wonderen$^e$,
Julea N. Butt$^c$, 
Hisao Ishii$^c$,
Israel Pecht$^f$, 
Mordechai Sheves$^f$, 
David Cahen$^f$, 
Jochen Blumberger$^{*,b}$} \\[1em]
\textit{
$^a$ University of South Bohemia, Faculty of Science, Branisovska 1760,\\ 370~05 Ceske Budejovice, Czech Republic \\
$^b$ University College London, Department of Physics and Astronomy, Gower Street,\\ London WC1E~6BT, UK \\
$^c$ Graduate School of Science and Engineering, Chiba University, Chiba, Japan \\
$^d$  Department of Materials and Interfaces, Weizmann Institute of Science, Rehovot, Israel \\ 
$^e$  School of Chemistry, School of Biological Sciences, University of East Anglia, Norwich Research Park, Norwich NR4~7TJ, UK \\
$^f$ Department of Organic Chemistry, Weizmann Institute of Science, Rehovot, Israel}
\end{center}

\begin{center}
\large{\textbf{Abstract}} \\
\normalsize
\vspace{1em}
\begin{minipage}{\textwidth}
Multi-heme cytochromes (MHC) are fascinating proteins used by bacterial organisms to shuttle electrons within and between their cells. 
When placed in a solid state electronic junction, they support temperature-independent currents over several nanometers that are three orders of magnitude higher compared to other redox proteins of comparable size. 
To gain microscopic insight into their astonishingly high conductivities, we present herein the first current-voltage calculations of its kind, for a MHC sandwiched between two Au(111) electrodes,  complemented by photo-emission spectroscopy experiments. 
We find that conduction proceeds via off-resonant coherent tunneling mediated by a large number of protein valence-band orbitals that are strongly delocalized over heme and protein residues, effectively ``gating" the current between the two electrodes.
This picture is profoundly different from the dominant electron hopping mechanism supported by the same protein in aqueous solution. 
Our results imply that current output in MHC junctions could be even further increased in the resonant regime, e.g. by application of a gate voltage, making these proteins extremely interesting for next-generation bionanoelectronic devices.
\end{minipage}
\end{center}
\vspace*{\fill}

\noindent \textbf{Author contributions:}\\[1ex]
Z.F. carried out the computer simulations, 
B.K., K.G. and I.P. performed the experimental I-V measurements, 
I.I. and H.I. performed the experimental UPS measurements,
X.J. modelled the I-V curves, 
J.H.v.W. and J.N.B. prepared the protein, 
Z.F. and J.B. analyzed the results, 
M.S, D.C. and J.B. designed the work.
\vspace{1em}

\noindent \textbf{Author declaration:}\\[1ex]
The authors declare no conflict of interest.
\vspace{1em}

\noindent \textbf{Keywords:}\\[1ex]
protein junction $|$ electron transport $|$ tunneling current $|$ heme $|$ density functional theory 

\clearpage


\section*{Introduction}

Redox-active metalloproteins are ubiquitous in living organisms facilitating many of the energy conversion processes that are quintessential for life on Earth including photosynthesis, respiration and nitrogen fixation. 
Recently, multi-heme cytochromes (MHC) and their complexes have gained much attention due to their involvement in extracellular respiration and inter-species electron exchange in dissimilatory metal-reducing bacteria~\cite{Breuer2015, Chong2018, Blumberger2018}. 
Atomic X-ray structures of several of these proteins were resolved for \emph{S. oneidensis}~\cite{Paixao2008, Clarke2011, Edwards2012, Edwards2015}, and most recently also for \emph{G. sulfurreducens}~\cite{Wang2019, Filman2019}, revealing closely packed heme c cofactor arrangements within the protein peptide matrices suggestive of their function as ``biological nanowires". 
Protein complexes~\cite{Hartshorne2009} (or polymers~\cite{Wang2019, Filman2019}) of MHCs span the entire bacterial envelope thereby facilitating the export of electrons from the inside to the outside of the cell. 
Experiments~\cite{vanWonderen2019, vanWonderen2018}, theory~\cite{Polizzi2012} and computation~\cite{Jiang2019, Jiang2017, Breuer2014, Breuer2012} have given valuable insights into the thermodynamics, kinetics and the mechanistic aspects of this process, in particular suggesting that electron transfer (ET) across these structures in their native (aqueous) environments occurs by consecutive heme-to-heme electron hopping~\cite{Jiang2019, Jiang2017, Blumberger2015, Breuer2014}.          
For nanotechnological applications, interfaces of the metalloproteins with solid electrodes are of great interest and their properties are intensively studied because of their potential utilization in enzymatic biofuel production, bioelectrocatalysis, biosensors and molecular (bio)electronics.~\cite{Szczesny2018, Bostick2018, Wu2007, Wong2003}
The ability to efficiently transport electrons is crucial for such applications and obviously some proteins are naturally designed to perform better than others. 
Recently, we demonstrated that MHCs are significantly better electronic conductors than other proteins in junctions composed of solid protein mono-layers in contact with two gold electrodes~\cite{Garg2018}. 
Most strikingly, the small tetraheme cytochrome (STC) was found to have a conductance three orders of magnitude higher than that of the blue copper protein Azurin~\cite{Sepunaru2011}, even though both proteins have similar cross section. 
Exceptionally large current densities (normalized to length) were also obtained for the deca-heme protein MtrF~\cite{Garg2018} that, were comparable with the STM single molecule currents reported earlier for MtrF~\cite{Byun2014}, MtrC~\cite{Wigginton2007a} and OmcA~\cite{Wigginton2007a} (see analysis in Ref.~\cite{Jiang2019}). 

Unfortunately, we currently still lack a good understanding of the atomistic origin of the large conductivities observed for MHCs, in contrast to our knowledge of their ET properties in aqueous solution~\cite{Blumberger2018}. 
This is highly unsatisfactory as it prevents us from rationally engineering multi-heme proteins for next-generation bionanoelectronics devices. 
While early single molecule STM measurements of MHCs were interpreted in terms of inelastic and elastic tunneling models~\cite{Wigginton2007b, Breuer2015}, more recent STM tunneling currents were modelled assuming activated heme-to-heme hopping similarly as for ET in solution~\cite{Pirbadian2012, Byun2014}. 
Latest measurements by Garg \textit{et al.} on MHC mono-layer junctions showed virtually zero temperature dependence between 320-80 K, pointing to tunneling as the dominant conduction mechanism over this temperature range.~\cite{Garg2018} However, this latter observation raises a number of questions: 
How can tunneling persist over such long distances in biological matter, that is, nanometer scales? 
What is the nature of the conduction channels that mediate the current? 
What is the role of the protein amino acids and the heme cofactors for conduction? 
Here we use theory and computation in combination with ultraviolet photoemission spectroscopy (UPS) to address these questions.  

While first principles calculations on molecular junctions (including peptides~\cite{Guo2016}) are now relatively standard, first-principles computations of the current-voltage ($I$-$V$) response for systems as large as proteins have so far not been reported, to our best knowledge. 
Here we calculate the $I$-$V$ curve for STC sandwiched between two Au(111) electrodes, treating all atoms of the system -- the full STC protein and the two gold electrodes -- at the density functional theory, more specifically DFT(PBE)+$\Sigma$ level of theory ($\approx 20,000$ electrons). 
The calculations are supported by UPS measurements which are used to determine the energy level alignment of protein states with respect to the Fermi-level of the electrodes. We also present the atomistic adsorption structures of STC on Au(111) as obtained from molecular docking and molecular dynamics (MD) simulation. 
We find strong evidence that electronic conduction is in the off-resonant coherent tunneling regime, mediated by a manifold of valence band orbitals that are delocalized over heme and protein amino acids. 
Fe is found to be unimportant for electronic conduction, whereas protein amino-acid residues in contact with the electrodes play a crucial role. 
Our findings suggest that the currents through STC could be increased even further if the energy levels of the protein are better aligned with the Fermi-level of the electrode, e.g. by application of a gate voltage.   

\section*{Results and Discussions}

\subsection*{Modelling of experimental I-V curves}

\begin{figure}
  \centering
  \begin{tabular}{l}
    \includegraphics[width=80.0mm]{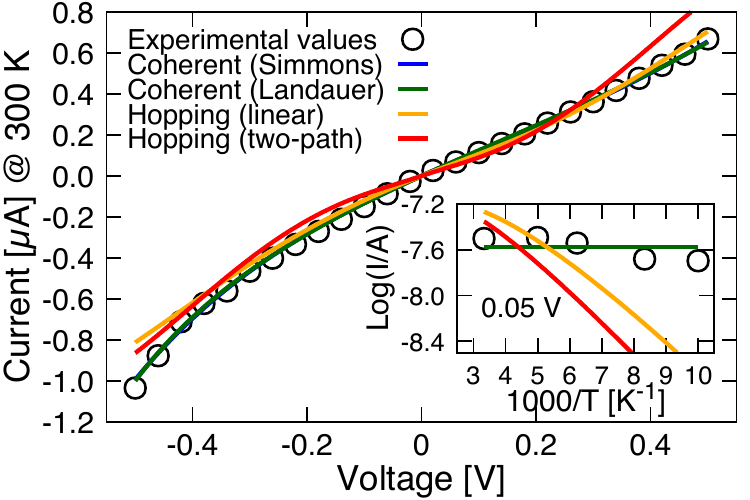}(a) \\
    \includegraphics[width=38.0mm]{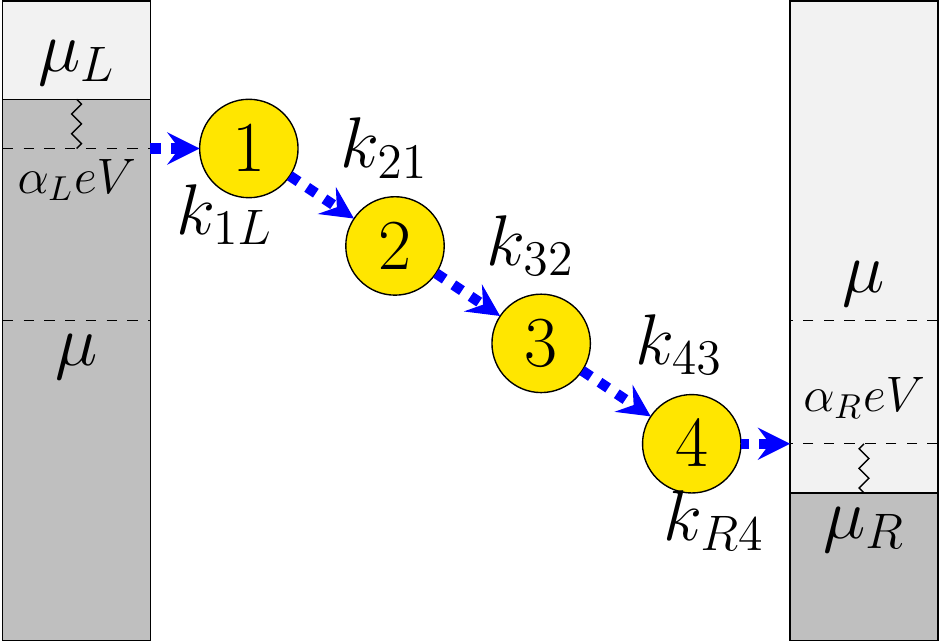}\,(b)
    \includegraphics[width=38.0mm]{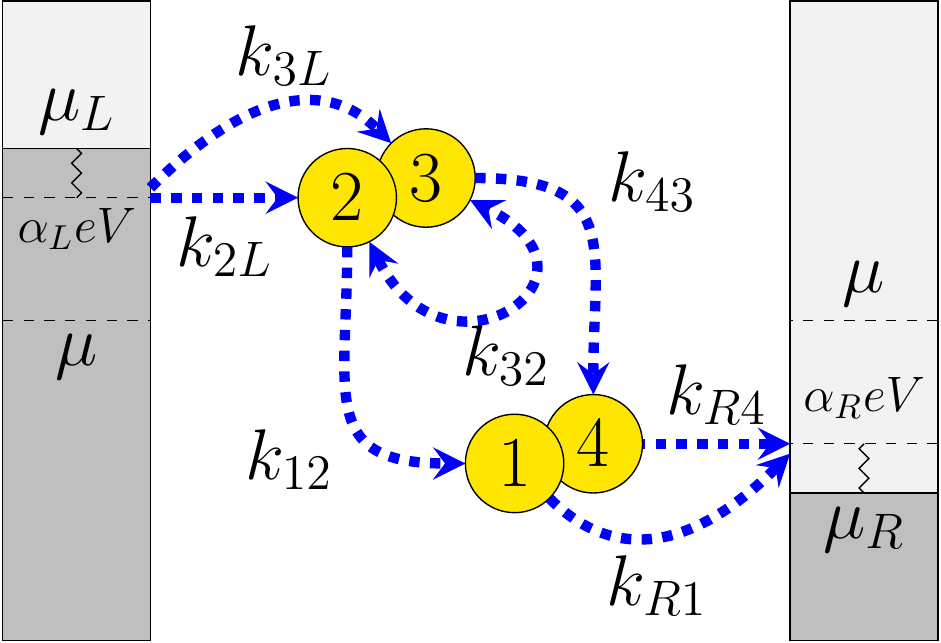}\,(c) \\
    \includegraphics[width=38.0mm]{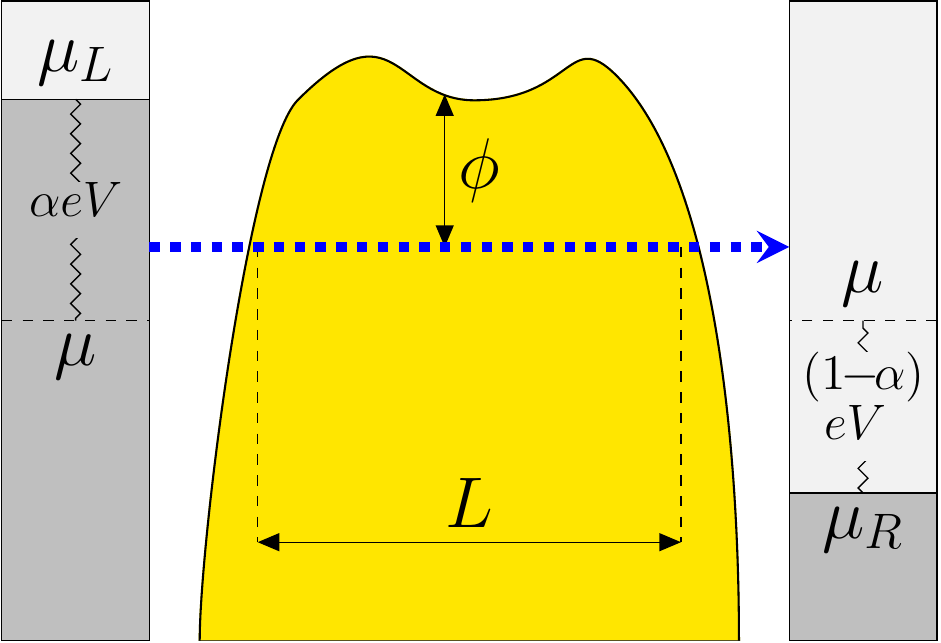}\,(d)
    \includegraphics[width=38.0mm]{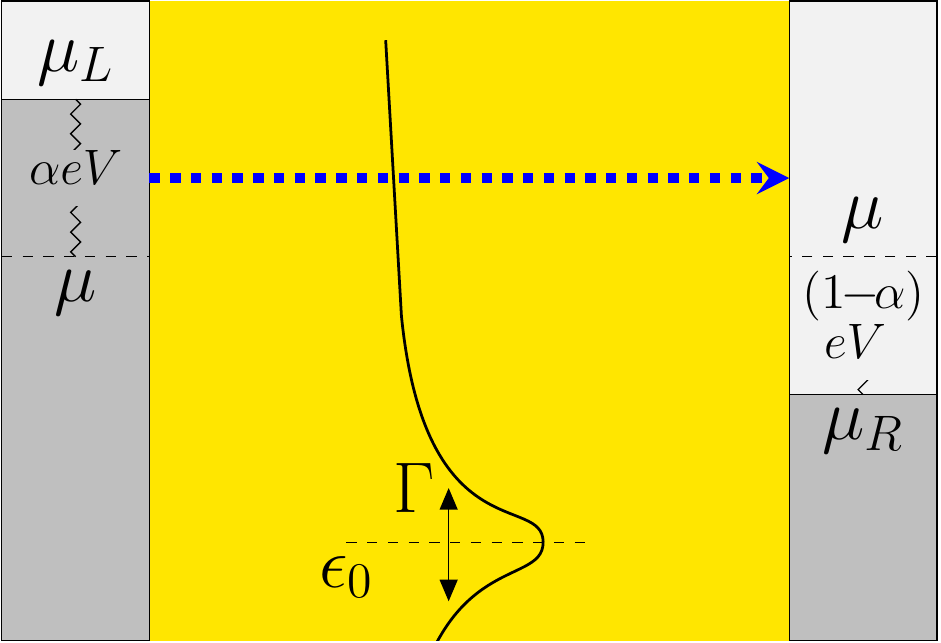}\,(e) \\
  \end{tabular}
  \caption{
(a) Current--voltage ($I$-$V$) and current--temperature ($I$-$T$) curves (inset) for the tetra-heme protein STC. 
The experimental data are obtained for a protein monolayer in vacuum using the suspended nanowire technique.\cite{Garg2018} 
Best fits of the experimental data are shown for incoherent hopping models (b) along a linear chain and (c) along a branched chain of heme cofactors, as well as for coherent tunneling models according to Simmons (d) and Landauer (e). 
Note that the coherent models of Simmons and Landauer predict the same fit and their curves are on top of each other.
$\mu_L$ and $\mu_R$ are the Fermi levels of the left (L) and right (R) electrode,  and $\alpha$ is the symmetry factor of the potential drop. 
In (b) and (c) $k_{ji}$ indicate the rate constant for ET from site $i$ to site $j$ and 1-4 denote the four heme cofactors. 
In (d) $L$ denotes the tunneling length and $\phi$ the tunneling barrier, and in (e) $\Gamma$ and $\epsilon_0$ are the width and position, respectively, of the effective conduction channel. 
See SI for a details.}
  \label{fig_1}
\end{figure}

Current-voltage ($I$-$V$) and current-temperature ($I$-$T$) curves are available for the tetra-heme protein STC with Cys introduced at site 87 that allows for chemisorption on a lower Au(111) surface.
We first analyze the available experimental data by fitting the measured curves to incoherent and coherent transport models (Fig.~\ref{fig_1}a) to show the qualitative differences between these two limit charge-transfer mechanisms.
It is well known that in aqueous solution the ET through STC occurs via incoherent heme-to-heme hopping. 
Hence, we investigate the same mechanism for the modelling of electronic conduction. 
The steady-state current at a given voltage was obtained by solving a chemical master equation for assuming the nearest-neighbour hopping model.~\cite{Breuer2014, Jiang2017, Polizzi2012}. 
The heme-heme hopping rate constants and the interfacial ET rates from/to the electrodes are calculated using the non-adiabatic Marcus expression~\cite{Marcus1985} and the Marcus-Hush-Chidsey expression~\cite{Chidsey1991}, respectively, where the voltage is assumed to modify only the ET free energies between the heme cofactors, that is their redox potentials.
Two possible adsorption geometries of the protein were investigated -- a 'standing' structure where the four-heme chain is aligned to the electrode-surface normal (Fig.~\ref{fig_1}b), as suggested in our earlier work~\cite{Garg2018}, and a 'lying' structure with two potential hopping pathways (see Fig.~\ref{fig_1}c). 
The latter structure is motivated by the adsorption geometry obtained from molecular docking of the protein on the gold surface discussed below. 
Further details on the incoherent hopping model and the fitting parameters are given in the Supporting Information (SI). 
Although it is possible to fit the $I$-$V$ and $I$-$T$ curves separately with two different sets of fit parameters, we find that neither of the two hopping models can capture both curves with a single set of fit parameters (Fig.~\ref{fig_1}a). 
A good fit of the $I$-$V$ curve (indicated in yellow and red in Fig.~\ref{fig_1}a for the two models) gives a too strong temperature dependence, whereas a good fit of the $I$-$T$ curve requires very small reorganization free energies of less than 0.1 eV and gives a qualitatively wrong shape of the $I$-$V$ curve.
 
In contrast to incoherent models, fully coherent electron tunnelling does not \emph{a priori} exhibit any temperature dependence (apart from that of the Fermi-Dirac distribution function) and therefore this mechanism seems to be more appropriate to explain the measured data in the STC junction. 
We first consider the popular Simmons model~\cite{Simmons1963} which assumes that the electron tunnels from one electrode to the other through a single potential barrier representing the protein region in the junction (Fig.~\ref{fig_1}d).
Indeed, this model can very well reproduce the experimental $I$-$V$ data ($R^2\!=\!0.9991$) predicting a barrier height $\phi\!=\!1.13$\,eV and an asymmetry coefficient of $\alpha\!=\!0.42$. 
However, the tunnelling distance $L\!=\!1.22$\,nm is much shorter than the experimental monolayer width of $2.4 \pm 0.5$\,nm as noted in our earlier work~\cite{Garg2018}.  
Yet we note that in Simmons' model the tunneling distance corresponds to the region with a barrier higher than that of the Fermi level of the negatively charged electrode and can therefore be shorter than the electrode distance.  

A more refined coherent tunneling model is provided by Landauer theory. 
Here, the structureless insulating material of the Simmons model is replaced by one or several molecular states mediating the charge transfer and effectively acting as so-called ``conduction channels" of the material (see Fig.~\ref{fig_1}e). 
\hl{For simplicity, we assume presence of one effective conduction channel of Lorentzian shape}~\cite{Carey2017, Valianti2019} \hl{for which we find that the Landauer model fits the experimental {$I$}-{$V$} curve almost perfectly} ($R^2\!=\!0.9995$).
The best fit gives a weak electrode-protein coupling $\Gamma\!=\!0.91$\,meV and an energy off-set for the conduction channel of $\epsilon_0\!=\!0.51$\,eV relative to the Fermi level at zero voltage. The weak coupling is in accord with a first survey of the protein surface indicating that the heme cofactors are not in direct contact with the gold electrodes and all amino acids at the interface except cysteine 87 are physi- rather than chemisorbed on the electrode. 
\hl{Note, that the generalized model with more conduction channels would provide the same agreement with the experimental curve, however, the large number of free parameters would lead to their ambiguous determinations due to the over-fitting of the relatively simple-shaped off-resonant experimental {$I$}-{$V$} curve.
Based on these analyses, we conclude that the conduction occurs via the coherent tunneling.
To investigate the system in more details, in particular to identify the molecular orbitals(s) mediating the current and explore the the structural features of the STC protein (Fe, heme, amino acids) that contribute to the current, electronic calculations on the atomistic model of the Gold-STC-Gold junction are required.}

\subsection*{Adsorption structure}

Since $I$-$V$ measurements were carried out in vacuum ($10^{-5}$ bar), we first verified using molecular dynamics (MD) simulation that the STC protein remains folded and stable under these conditions. 
Indeed, the RMSD remained rather small, 1.5\,\AA\ with respect to the crystal structure, along a trajectory of length 40 ns. 
The robust secondary structure is a result of strong covalent binding of the rigid heme cofactors to the protein matrix via cysteine linkages and of axial coordination of the heme iron cations to two proximal histidines. 
In addition, three structural water molecules were found trapped inside the protein by a strong hydrogen-bond network, which helps keep the exposed and more-flexible protein loops in a folded configuration.
 
The experiments were carried out for the serine 87-to-cysteine (S87C) mutant of STC, allowing the sulfur atom of this cysteine to bind covalently to the bottom gold contact. 
The polycrystalline thin films of Au, have predominantly the (111) orientation, as we checked for our films by X-ray powder diffraction (XRD).
Hence, \emph{in silico} we placed the vacuum-relaxed S87C mutant of STC 10~\AA\ away a Au(111) surface slab.
We used the GolP-CHARMM force field~\cite{Iori2009, Wright2013} to model the interactions between gold and the protein. 
This force field takes into account image charge effects and gives fairly good adsorption structures and energies when compared with accurate van der Waals density functional (vdW-DF) calculations~\cite{Dion2004, Futera2019}. 
Carrying out MD simulations at a temperature of 300~K we find that the protein spontaneously adsorbs on the gold surface. 
We repeated the MD simulation runs for 144 different initial protein structures differing in orientation of the protein with respect to the surface, of which 49 \% adsorbed on the gold surface by Cys-87 site.

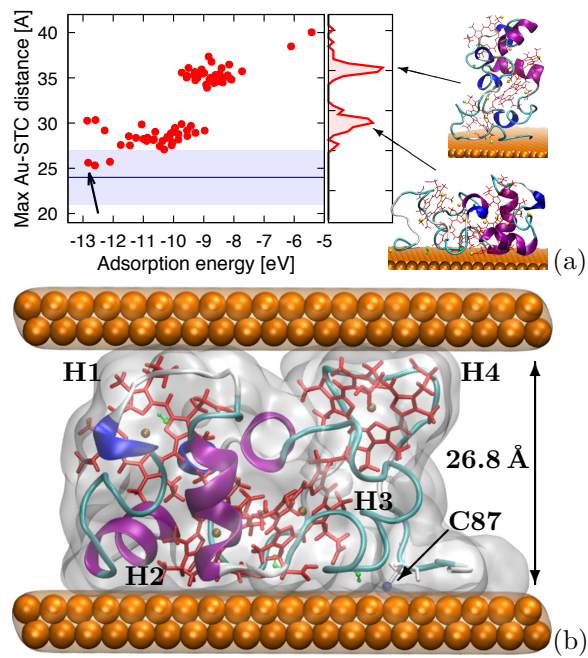
\begin{figure}
  \centering
  \begin{picture}(217,260)
    \put(  0,150){\includegraphics[scale=0.9]{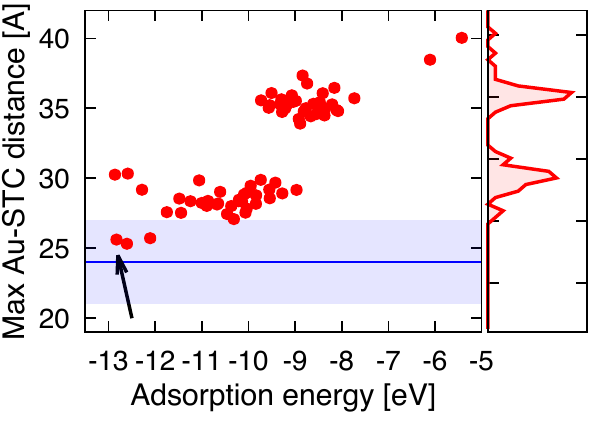}}
    \put(175,200){\includegraphics[scale=0.8]{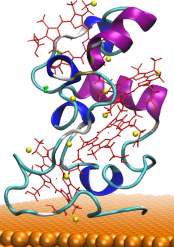}}
    \put(151,155){\includegraphics[scale=0.9]{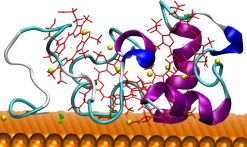}}
    \put(217,155){(a)}
    \put(180,230){\vector(-4, 1){25}}
    \put(170,195){\vector(-3, 2){25}}
    \put(  0,  0){\includegraphics[scale=0.9]{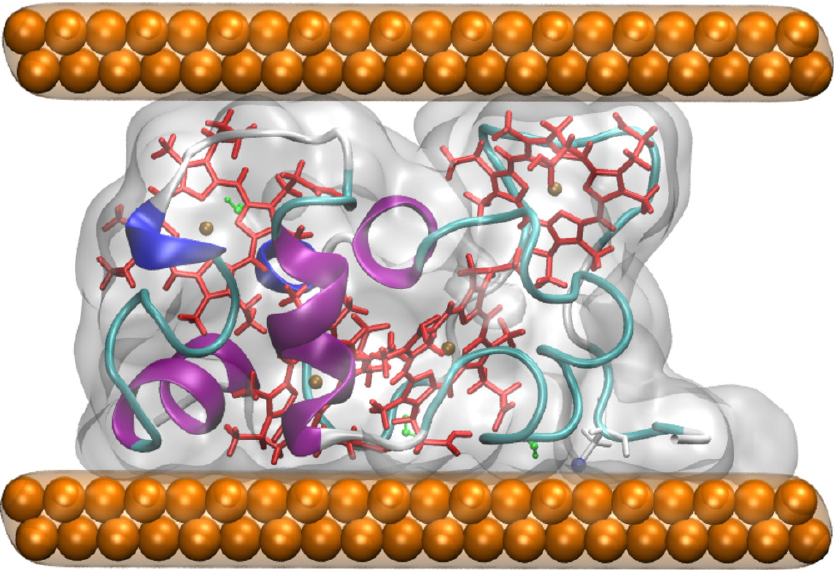}}
    \put(217,  2){(b)}
    \put( 20,110){\textbf{\color{black}{H1}}}
    \put( 45, 28){\textbf{\color{black}{H2}}}
    \put(137, 58){\textbf{\color{black}{H3}}}
    \put(180,110){\textbf{\color{black}{H4}}}
    \thicklines
    \put(210, 28){\vector( 0, 1){90}}
    \put(210,118){\vector( 0,-1){90}}
    \put(174, 75){\textbf{26.8\,\AA}}
    \put(175, 50){\vector(-1,-1){20}}
    \put(175, 50){\textbf{C87}}
  \end{picture}
  \caption{
Adsorption structures of STC on Au(111) as obtained from docking and MD simulation. In (a) the largest distance of any protein atom in the surface-normal direction is plotted against adsorption energy for generated samples.
The structures can be clustered in two distinct adsorption geometries, 'lying' and 'standing' (red distributions with pointed representative structures). 
The experimental range and mean value of mono-layer thickness is indicated in (a) by shaded area and the blue line, respectively.
In (b) the 'lying' structure indicated by an arrow in panel (a) is shown after the top electrode contact is added.
The protein is chemisorbed to the bottom contact via Cys-87 and physisorbed to the top electrode. 
The heme cofactors are show in red and secondary structure elements of STC in cartoon representation.}
  \label{fig_2}
\end{figure}

The adsorption structures generated could be clustered in two groups -- the 'standing' configuration where the heme chain is orthogonal to the gold surface, and the 'lying' configuration where the heme-chain is parallel to the surface (Fig~\ref{fig_2}a). 
As one may expect, the latter structures have a significantly larger adsorption energy, by about 3~eV, and are thus more stably attached to the surface. 
Moreover, only the lying structures are within the experimental range $2.4\!\pm\!0.5$~nm of mono-layer thickness~\cite{Garg2018} (blue area in Fig~\ref{fig_2}a). 
Therefore, we chose the horizontal structure with the smallest RMSD (2.3~\AA\ compared to STC crystal structure; indicated by an arrow in Fig~\ref{fig_2}a), which also turned out to have one of the highest adsorption energies. 
This structure was chemisorbed to the surface by specifying a covalent interaction between the sulfur atom of Cys-87 and gold using the Au-S covalent interaction parameters fitted previously to vdW-DF calculations. 
To complete the structural model of the junction, the top contact was placed at close contact with the upper protein surface. 
After protein relaxation, the distance between the two electrodes was varied until the local pressure tensor in the protein region integrated to zero.~\cite{Thompson2009, Ollila2009} 
The final electrode separation obtained was 2.7\,nm in good agreement with experimental measurements, $2.4 \pm 0.5$\,nm.~\cite{Garg2018}
In the final protein structure (shown in Fig.~\ref{fig_2}b) hemes 2 and 3 are in proximity with the bottom contact, whereas hemes 1 and 4 are close to the upper contact, thus forming a bifurcated heme path between the electrodes (as mentioned above in {\it Modelling of experimental $I$-$V$ curves}).
This structure was used for electronic structure and $I$-$V$ calculations at all-QM DFT level, as detailed below.

\subsection*{Electronic band structure}

The electronic structure calculations on the full Gold-STC-Gold model junction were performed with the CP2K software package using the PBE functional, GTH pseudopotentials and the DZVP basis set~\cite{Hutter2014, Goedecker1996, Perdew1996}.
Although this functional can describe metallic states of gold rather well, it suffers from an inaccurate band alignment of the protein states with respect to the Fermi level of the electrode. 
The band alignment is important for the conductive properties of the protein and therefore we used the DFT+$\Sigma$~\cite{Neaton2006, Quek2007, Darancet2012, Liu2014, Egger2015} approach to correct it (see SI for details). 
We shift the occupied protein states by 1.2~eV downwards placing the highest occupied molecular orbital (HOMO) at $-1.2$ eV with respect to the Fermi level, and the unoccupied states by 1.4~eV upwards placing the lowest unoccupied molecular orbital (LUMO) at 1.4~eV. 
Although one cannot expect DFT+$\Sigma$ to be quantitative in general\cite{Liu2017}, we will see below that the predicted HOMO alignment is in good agreement with ultraviolet photoelectron spectroscopy (UPS) data.   

\begin{figure}
  \centering
  \begin{tabular}{rl}
    \setlength{\tabcolsep}{0pt}
    \includegraphics[width=70mm,height=35mm]{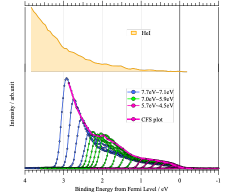} & \hspace{-5mm}(a) \\
    \includegraphics[width=70mm,height=35mm]{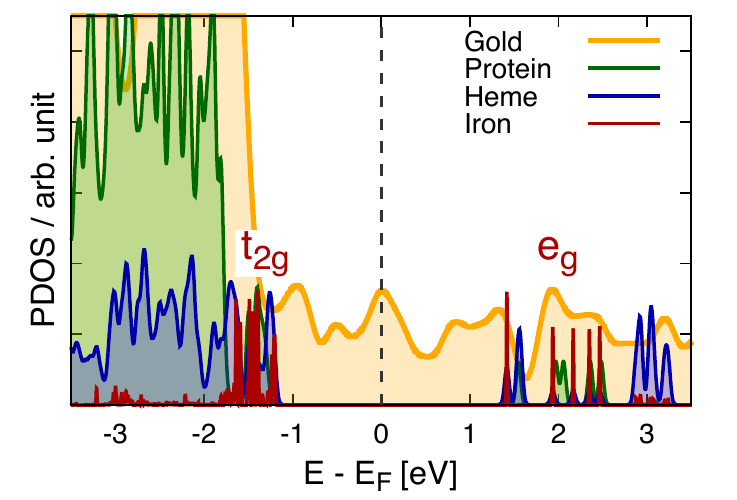} & \hspace{-5mm}(b)
  \end{tabular}
  \caption{
(a) \hl{UPS signal of STC monolayer on Au substrate with photon excitation energy of HeI (21.22~eV; top) and its resolution with variable low energy UV light ({$h\nu$} = 4.5 –- 7.7~eV in 0.2~eV steps).
The envelope of the 4.5 –- 7.7~eV spectra shows  a small peak structure around 2~eV while the HeI spectrum shows only a monotonic tail (towards 0~eV, the Fermi level). 
Constant final state yield (CFS) plot at {$E_k$} = 0.3~eV superimposed on the measured low energy UPS signals. 
The density-of-state curve is estimated from the CFS plot and thus shows a protein onset energy of {$\sim$}1.2~eV (intersection of the two straight lines, drawn on the CFS plot, one line for the Au sp levels and one for the protein levels.}
(b) Projected density of states (PDOS) of STC near the Fermi level $E_F$. 
The states were obtained from Kohn-Sham DFT calculations and localized on protein and gold electrodes using the projector operator-based diabatization method (POD). 
The orbital energy of the the POD states were shifted using the DFT+$\Sigma$ approach (see main text and SI for details).}
  \label{fig_3}
\end{figure}

For experimental determination of the protein valence band edge and the Fermi level of the Au substrate ($E_F$, work function), the STC protein was adsorbed on the Au substrate, in the same way as was done for the $I$-$V$ measurements.
Fig.~\ref{fig_3}a shows the UPS signals, obtained with a Helium discharge lamp (HeI: $h\nu\!=\!21.22$~eV) as well as the signal obtained with monochromatized low, variable energy UV ($h\nu\!=\!4.5-7.7$~eV) excitation.
Because of the very low contribution of stray light in the experimental system that is used, the background signal is significantly suppressed, enabling high-sensitivity measurement~\cite{Sato2017, Machida2013}. 
In the HeI spectrum, only a tail structure that decays monotonically is observed, probably due to slight sample charging and the small signal intensity from the STC's HOMO state. 
In contrast, in the low energy UV spectra a peak structure around 2 eV develops as the UV photon energy is varied from 7 to 5.9~eV.
For $h\nu < 5.7$~eV, no special features are seen in the peak sequence. 
These spectral intensities basically reflect the density-of-state (DOS) of the film weighted by the $h\nu$-dependent photoionization cross-section.

To reduce the cross-section factor, constant final state (CFS) yield spectroscopy was proposed~\cite{Korte2008}, in which the intensity of photoelectrons with a constant kinetic energy ($E_k$) is recorded as function of incident photon energy to cancel the final state effect on the cross-section. 
From the CFS-plot, the Fermi edge and the following sp-band structure of the Au substrate appear up to 2 eV. 
The structure from 2.4 to 3.2~eV can be ascribed to an overlap of photoemission from the d band of Au and from protein states. 
A peak structure around 2~eV is observed. By using a linear extrapolation we get the energy of the protein HOMO onset, $1.2$~eV below the Fermi level as predicted by DFT+$\Sigma$.

The final electronic band structure of the Gold-STC-Gold junction as obtained from DFT+$\Sigma$ calculations is shown in Fig~\ref{fig_3}b. 
The total projected density of states is broken down in contributions from Fe atoms (denoted ``Iron"), the porphyrine rings and axial histidines ligating Fe (collectively denoted ``Heme"), all amino acid residues except the axial histidines (denoted ``Protein") and gold. 
We find that the highest valence band states of the protein gives rise to three distinct peaks between $-1.2$ and $-1.8$~eV and correspond to Fe d $t_{2g}$ states hybridized with orbitals from the porphyrine ring, the axial histidines and partly also on the cysteine linkages of the heme cofactors.
Some of these states are localized on a single heme while others are delocalized over up to all 4 heme groups of STC.
The iron band is mixed with the highest protein amino-acid electronic states localized on Met-67 (-1.3~eV) in the middle of the junction, Asp-81 (-1.4~eV) near the upper gold surface, and N-terminal acetyl (-1.4~eV), Ser-37 (-1.4~eV) and Gly-70 (-1.5~eV), which are amino acids physisorbed on the bottom gold surface.
At energies lower than -1.8~eV the PDOS for heme and amino acids strongly increases without any significant contributions from Fe. 
The lowest conduction band between $1.4$ and $2.5$~eV is comprised of the Fe d $e_g$ states mixed with heme and, to a smaller extent, protein amino-acid orbitals. 
The latter are due to five lysines that are strongly physisorbed to the bottom Au surface in the vicinity of chemisorbed Cys-87 (Lys-72 (1.5~eV), Lys-90 (2.0~eV), Lys-91 (2.1~eV), Lys-41 (2.4~eV) and Lys-50 (2.5~eV). 
The PDOS of gold is very similar to the one for the bare gold surface with the typical 3d band edge at -1.5~eV and a wide band of 4s states near and above the Fermi level. 
Having characterized the PDOS, we are now in a position to calculate the Landauer current and to interpret the measured electronic transport behaviour of STC at an atomistic level of detail.   

\subsection*{Computation of $I$-$V$ response}

In the Landauer-B\"uttiker formalism, the tunnelling current, $I$, is obtained as an integral of the transmission function over the Fermi window for a given applied voltage, $V$, 
\begin{equation}
  I(V) = \frac{e}{\pi\hbar} 
    \int T(E) \left[ f_L(E,V) - f_R(E,V) \right] dE 
  \label{eq_iv_landauer}
\end{equation}
where $E$ is the energy of the tunneling electron and $f_M$, $M = \{L, R\}$ are the Fermi-Dirac distributions on the left (L) and right (R) electrode, respectively.

Here, we use \hl{the Breit-Wigner approximation of the transmission function}~\cite{Carey2017, Papp2019} \hl{where each molecular state forms the Lorentzian peak of the width given by its coupling to the left and right electrodes, respectively.}
The transmission is then sum of such peaks
\begin{equation}
  T(E) = \sum_{j}^{\mathrm{protein}}  \frac{\Gamma^{(L)}_j \Gamma^{(R)}_j}{
    [E - \epsilon_j]^2 + \Gamma_j^2}, 
  \qquad
  \Gamma_j = (\Gamma^{(L)}_j +\Gamma^{(R)}_j)/2  
  \label{eq_transmission}
\end{equation}
where $\Gamma^{(M)}_j$ are the spectral density functions determining the width of the Lorentzian $j$, and  $\epsilon_j$ is the energy of the protein molecular orbital $j$. 
The spectral density functions are proportional to the electronic coupling between protein MO $j$ and electrode states $q$ at given energy level, $\Gamma^{(M)}_j\!=\!2\pi\sum_q^M |H_{q,j}|^2\delta (E-E_q)$. 
We applied the projection-based diabatization (POD) method to calculate the electronic coupling elements $H_{q,j}$ and the one-particle protein and electrode energy levels, $\epsilon_j$ and $E_q$, respectively~\cite{Futera2017}. 
See SI for further details. 

\begin{figure}
  \centering
  \includegraphics[scale=0.8]{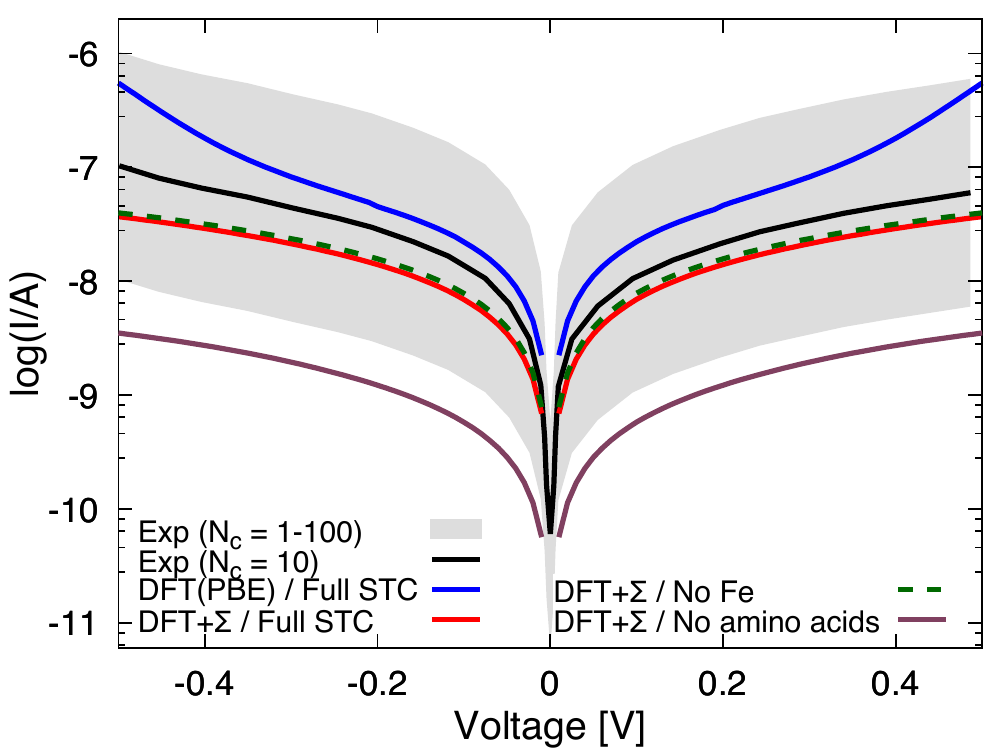}
  \caption{
Calculated $I$-$V$ curves for the Au-STC-Au junction shown in Fig.~\ref{fig_2}b.
The currents are computed within the Landauer formalism Eq.~\ref{eq_iv_landauer}
 using all-QM calculations on the entire junction, specifically projection operator-based diabatization (POD) in combination with DFT(PBE)+$\Sigma$ (red line). 
The estimated experimental current per STC protein is shown in black lines. 
Assuming that the device contains 10 active protein contacts, the current per protein was obtained by dividing the as-measured current shown in Fig.~\ref{fig_1}a by a factor of 10. 
The likely error bar for this estimate is shown in shaded gray corresponding to 1-100 active protein contacts in the device. 
The $I$-$V$ curve obtained with DFT(PBE), i.e. without $\Sigma$ correction, is shown in blue.
$I$-$V$ curves for modified STC structures with Fe atoms replaced by 2 H atoms (green dashed) and without protein amino acids(purple) are shown for comparison.}
  \label{fig_4}
\end{figure}

The results of the current calculations are summarized in Fig.~\ref{fig_4}. 
We find that the computed $I$-$V$ curve (red line) is in excellent agreement with the experimental data (black line), matching both the shape and the magnitude of the current response
 and justifying thus the applied approximation of the transmission function.
Importantly, the transport is in the off-resonant regime because all occupied protein states are at energies lower than $-1.2$\,eV and outside the Fermi window opened by the experimental voltage range (0.5 V) and so are the unoccupied states.
Hence, the current increases smoothly with voltage and does not contain any resonant peaks. 
As the transmission function is flat in the Fermi window the current response to the applied voltage is practically linear, that is Ohmic. 
To explore how the current-voltage curve would look like in the resonant regime, we shift all occupied protein levels upwards by 1.2 eV so that the protein HOMO is aligned with the Fermi-levels of the electrode at zero voltage. 
The shape of the $I$-$V$ curve is now rather different (Fig.~\ref{fig_4}, blue line). 
The resonant molecular states give rise to a rapid increase in the current for small voltages, as one would expect, and there is another stronger increase at about 0.4 V. 
These non-Ohmic responses, induced by the shifted valence band peaks of the protein within the Fermi window, are not seen in experiment and a further confirmation that electron transport is indeed in the off-resonant tunneling regime.

\subsection*{Conduction channels}

\begin{figure*}
  \centering
  \begin{picture}(480,190)
    \put(  0, 95){\includegraphics[width=230pt,height=90pt]{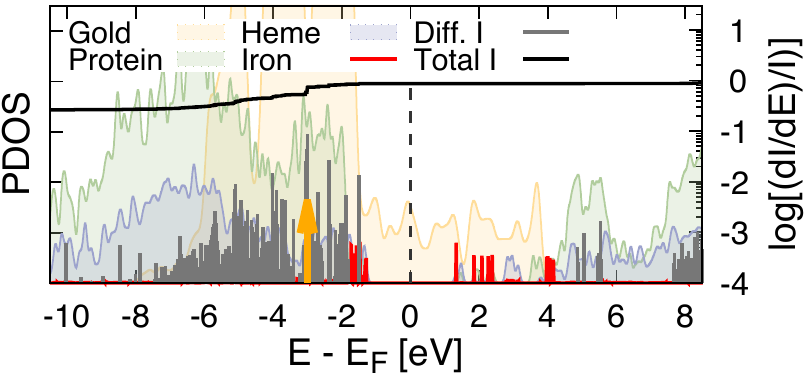}}
    \put(240, 95){\includegraphics[width=230pt]{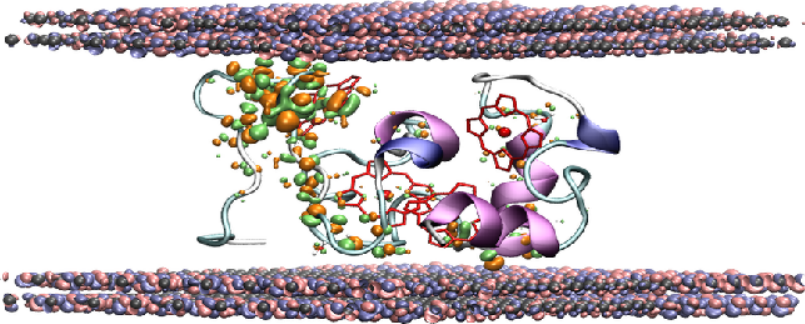}}
    \put(  0,  0){\includegraphics[width=230pt,height=90pt]{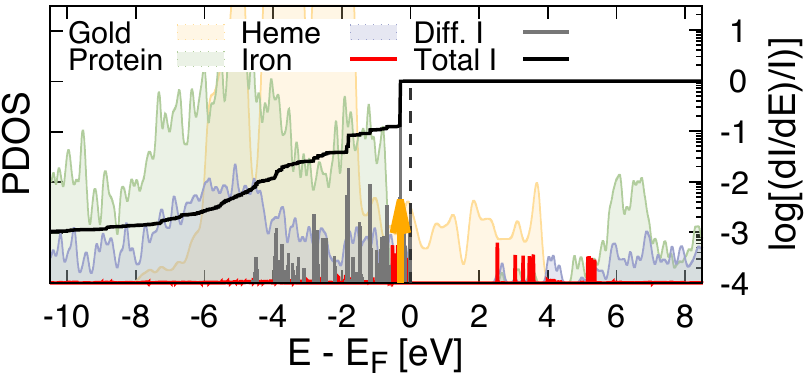}}
    \put(240,  0){\includegraphics[width=230pt]{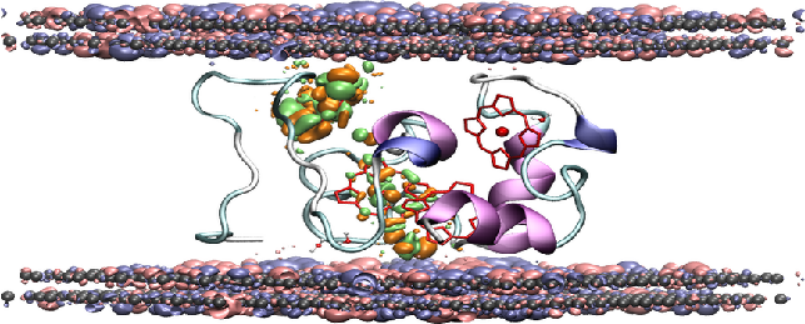}}
    \put(225, 97){(a)}
    \put(470, 97){(b)}
    \put(225,  2){(c)}
    \put(470,  2){(d)}
  \end{picture}
  \caption{
Breakdown of the total current in contributions from molecular orbitals of the STC protein. 
In (a) the differential current contributions $\log{[(dI/dE)/I]}$ (grey) to the total current $\int_{-\infty}^E (dI/dE')/I dE'$ (black) are shown for all molecular orbitals.
The orbitals are the same as the ones used for the current calculation in Fig. 4 (i.e. obtained from POD calculation in combination with DFT(PBE)+$\Sigma$) and are shown relative to Fermi level of the electrodes at zero voltage.
The corresponding projected density of states (PDOS) is shown as well and broken down in contributions from gold, protein amino acids, heme and iron. 
In (b) the molecular orbital of STC with the highest contribution to the current (marked by an orange arrow in panel (a)) is depicted in orange and green isosurfaces. 
The two metallic states in the bottom and top electrodes exhibiting the largest coupling with this molecular orbital ($\Gamma^{(L)}$ and $\Gamma^{(R)}$) are depicted in blue and pink isosurfaces.
The analogous data for the (hypothetical) resonant tunneling regime where the molecular orbitals are shifted upwards by 1.2~eV are shown in (c) and (d).}
  \label{fig_5}
\end{figure*}

Which protein states mediate the tunneling current? 
To answer this question we plot in Fig.~\ref{fig_5}a the contribution of each protein state to the total current at a voltage of 0.5~V (gray) as well as the accumulated sum (black line). 
The corresponding projected density of states is shown as well.
We find that the current is the result of many small contributions originating from protein and heme states at energies between about -8 and -2 eV. 
These states are typically delocalized over two or three heme cofactors that bridge the two electrodes (0.20--0.25\%) and the protein amino acids (0.80--0.75\%). 
As a representative example, we show the molecular orbital of STC with the largest contribution to the total current, 8.8\% (positioned at -3.0~eV), in Fig.~\ref{fig_5}b. 
Interestingly, the highest valence band of the protein composed of the Fe d $t_{2g}$-heme orbitals contribute very little to the current, even though these states are closest in energy to the Fermi window. 
The reason is that their coupling to the electrode ($\Gamma^{(L)}$ and $\Gamma^{(R)}$) is much smaller than for the most conductive states since they are mostly localized on the heme and do not spread over the amino acids that are in van-der Waals contact with the electrodes. 
Unoccupied states up to 10 eV above the Fermi level were involved in the calculations, however, their contribution to the tunnelling current is negligible (see Fig.~\ref{fig_5}a).
In particular, Fe $e_g$-heme orbitals located at the conduction band edge do not affect the currents and the conductivity is mediated predominantly by the valence-band states.

The situation is strikingly different for the (hypothetical) resonant regime considered before where all protein states are shifted upwards by 1.2 eV to align with the Fermi level at zero voltage. 
The current contributions and projected density of states are shown in Fig.~\ref{fig_5}c. 
In this regime, 83\% of the current is due to the highest valence band comprised of Fe d $t_{2g}$-heme orbitals typically delocalized over 2-3 hemes, 
see Fig.~\ref{fig_5}d for the orbital with the highest contribution (positioned at $-0.3$~eV after the shift is applied). 
The reason is that the transmission function Eq.~\ref{eq_transmission} for these near-resonant levels ($r$) becomes 
$T(E\!\approx\!\epsilon_r)\!\approx\!4 \Gamma^{(L)} \Gamma^{(R)}/(\Gamma^{(L)} + \Gamma^{(R)})^2$, which is maximum, $T\!=\!1$, for $\Gamma^{(L)}\!=\!\Gamma^{(R)}$. 
Hence, for near-resonant states to have large transmission the absolute value of the couplings to the electrode can be small, as is the case for 
Fe d $t_{2g}$-heme orbitals, as long as the couplings to both electrodes is symmetric. 
The conduction channel shown in Fig.~\ref{fig_5}c is delocalized mostly on the 
first and second hemes (H1 and H2 in Fig.~\ref{fig_2}b) and partly on H3, hence forming an ideal connection between the two electrodes.
This results in only modest asymmetry in the coupling values compared to most other conduction channels ($\Gamma^{(L)}\!=\!13.4$~meV and $\Gamma^{(R)}\!=\!2.0$~meV) and a relatively broad transmission peak of significant height ($T$ = 0.45).

\subsection*{Importance of Fe and protein to conductance}

To further understand the role of Fe, heme cofactors and protein amino acids in determining conductance of STC, we calculated the current-voltage curve for two different protein modifications, all based on the same Gold-STC-Gold structural model used before: 
(i) the Fe atom in each heme is replaced by 2 H atoms,
(ii) all protein amino acids are removed retaining only the Fe-heme cofactors, axial histidines and cysteine linkages. 
We find that replacement of Fe has virtually no effect on the current-voltage response (see red vs. dashed green line in Fig.~\ref{fig_4}). 
By contrast, removal of all protein amino acids leads to a significant drop in the tunneling current by one order of magnitude (purple line). 
Considering the analysis of the conduction channels in the unmodified STC protein, this result is not unexpected. 
It shows that in the present off-resonant regime, most of the coupling with the electrodes is due to protein amino acids and that the mixing of the protein states with the Fe-heme states is not essential. 
The insignificant role of Fe for conduction is in line with previous experimental measurements of conductance in Fe-containing and Fe-free cytochrome c.\cite{Amdursky2013} 

\section*{Concluding remarks}

Combining temperature-dependent conductance measurements, photo-emission spectroscopy (UPS) and large-scale DFT calculations, we have uncovered the conduction mechanism through solid state multi-heme protein junctions.   
The data unequivocally rule out activated hopping and strongly suggest off-resonant coherent tunneling over $\sim$~3~nanometers as the dominating conduction mechanism. 
DFT calculations within the Landauer formalism show that the active transport channels (i.e. MOs of STC) are delocalized over typically 2-3 hemes and strongly mix with orbitals of amino acid residues that are in van-der-Waals contact with the electrodes. 
We find that the total current is a collective effect of a few hundred of such states, each contributing a small fraction. 
The reason for this is that the valence band edge of STC is rather deep in energy in the monolayer junctions ($\le -1.2$ eV with respect to the Fermi energy of the electrodes) giving rise to a flat transmission function in the Fermi-window for each conduction channel. 
The same picture may explain previous single molecule STM measurements for the deca-heme proteins MtrC\cite{Wigginton2007a} and MtrF\cite{Byun2014}. 
However, the partial protein solvation and the possibly different energy level alignment in those measurements might tip the balance between this and other mechanisms. 

Our findings imply that the mechanism for electronic conduction through solid MHCs mono-layers in vacuum is fundamentally different from (chemically induced) electron transfer across the same protein in aqueous solution. 
While the latter proceeds via consecutive Fe$^{2+/3+}$ hopping mediated by the redox-active Fe d($t_{2g}$)-heme orbitals at the top of the valence band, conduction occurs by a manifold of valence band states delocalized over heme and protein amino acids. 
The role of the protein matrix is to augment the tails of the heme orbitals to increase the electronic coupling with the electrodes: without such contributions to the coupling, the protein conduction sharply decreases because the heme edges cannot fully approach the electrodes due to steric hindrance. 
Our results thus provide now an explanation for the earlier experimental finding that conduction through cytochrome c does not require Fe~\cite{Amdursky2013} whereas iron is mandatory for electron transfer redox activity. 
Still, in STC Fe has an important structural role keeping the protein rigid and preventing unfolding of the main secondary structure motifs when the protein adsorbs on the metal surface, according to our MD simulations.  
 
Intriguingly, the conduction mechanism changes qualitatively in the resonant regime where the protein valence-band edge is aligned with the Fermi energy as reported recently for cytochrome c~\cite{Fereiro2019} and earlier for azurin~\cite{Fereiro2018}.
In this scenario, the electron transport is dominated by the familiar Fe d($t_{2g}$)-heme orbitals that mediate electron hopping in solution, more specifically by linear combinations thereof with contributions on 2-3 hemes that bridge the space between the electrodes. 
Hence, by tuning the energy onset between the protein states and the electrode work function, which can in principle be done by suitable protein mutations, surface modifications or application of a gating potential, as reported recently for azurin~\cite{Kayser2018}, one can control the active states for electron transport. 
Although such modifications might be, and certainly solid state gating still is, non-trivial in practice, knowledge of the electronic states and their positions provides useful guidance for control and design of bioelectronic devices.

\section*{Acknowledgements}

Z.F. was supported by EPSRC (EP/M001946/1) and by the European Research Council (ERC) under the European Union's Horizon 2020 research and innovation programme (grant agreement no. 682539/ SOFTCHARGE) and by the Czech Science Foundation (project 20-02067Y).
X.J. was supported by a Ph.D. studentship cosponsored by the Chinese Scholarship Council and University College London. 
J.H.v.W. was supported by EPSRC (EP/M001989/1).
I.I. and H.I were supported by JSPS KAKENHI Grant Number (16H04222).
MS and DC thank the Israel Science Foundation and the German Science Foundation (DFG) for partial support.
Via UCL-group membership of the UK's HEC Materials Chemistry Consortium, which is funded by EPSRC (EP/L000202, EP/R029431), this work used the ARCHER UK National Supercomputing Service (http://www.archer.ac.uk). 
We are grateful to the UK Materials and Molecular Modelling Hub for computational resources, which is partially funded by EPSRC (EP/P020194/1).

\section*{Supporting Information}

Models for coherent and incoherent electron transport, details of the classical atomistic simulations, tunnelling current calculations based on the Landauer formalism and the POD method, electronic band alignment, localization of the conduction channels.


\small{
\bibliography{references.bib}
\bibliographystyle{references}
}

\clearpage

\clearpage

\clearpage

\clearpage

\clearpage

\end{document}


\begin{center}
\Large{
Off-resonant Coherent Electron Transport over Three Nanometers in Multi-heme Protein Bioelectronic Junctions} \\[1em]
\normalsize{
Zdenek Futera$^{a,b}$, 
Ichiro Ide$^c$,
Ben Kayser$^d$
Kavita Garg$^d$
Xiuyun Jiang$^b$, 
Jessica H. van Wonderen$^e$,
Julea N. Butt$^c$, 
Hisao Ishii$^c$,
Israel Pecht$^f$, 
Mordechai Sheves$^f$, 
David Cahen$^f$, 
Jochen Blumberger$^{*,b}$} \\[1em]
\textit{
$^a$ University of South Bohemia, Faculty of Science, Branisovska 1760,\\ 370~05 Ceske Budejovice, Czech Republic \\
$^b$ University College London, Department of Physics and Astronomy, Gower Street,\\ London WC1E~6BT, UK \\
$^c$ Graduate School of Science and Engineering, Chiba University, Chiba, Japan \\
$^d$  Department of Materials and Interfaces, Weizmann Institute of Science, Rehovot, Israel \\ 
$^e$  School of Chemistry, School of Biological Sciences, University of East Anglia, Norwich Research Park, Norwich NR4~7TJ, UK \\
$^f$ Department of Organic Chemistry, Weizmann Institute of Science, Rehovot, Israel}
\end{center}

\vspace{-0.9cm}
\begin{center}
$^*$ to whom correspondence should be addressed:
     e-mail: j.blumberger@ucl.ac.uk (J.B.) 
\end{center}     

\begin{center}
\Large{\textbf{Supporting Information}}
\end{center}

\clearpage


\section{Modelling of experimental $I$-$V$ curves}

\subsection{Incoherent hopping model}

Sequential electron transfer from one electrode to the other via a series of intermediate redox states is assumed to happen incoherently when full relaxation of the charge on one site is assumed before it moves on to the next site. Fully incoherent electron transfer rates from site $j$ to site $i$ are determined by the non-adi\-aba\-tic Marcus formula,~\cite{Marcus1985}
\begin{equation}\label{eqn_k_marcus}
  k_{ij} = 
    \frac{2\pi}{\hbar} 
    \left\langle|H_{ij}|^2\right\rangle
    \frac{1}{\sqrt{4\pi\lambda_{ij}k_BT}}
    \exp{\left(-\frac{(\lambda_{ij}+\Delta G_{ij})^2}{4\lambda_{ij}k_BT}\right)},
\end{equation}
where $H_{ij}$ are the electronic coupling elements between the two sites, $\lambda_{ij}$ is the reorganization free energy, $\Delta G_{ij}$ is the driving force (redox-potential difference between sites $i$ and $j$ at given voltage) and $k_BT$ is the thermal energy.
On the other hand, electrochemical electron transfer rates between an electrode and a redox site is given by the Marcus--Hush--Chidsey integral formula~\cite{Chidsey1991, Polizzi2012}
\begin{subequations}\label{eqn_k_chidsey}
\begin{equation}
  k_{iM} = 
    \frac{\Gamma}{\hbar} \sqrt{\frac{k_BT}{4\pi\lambda_i}} \int 
    \exp{\left[-\left(x-\frac{\lambda_i + \mu_M - eE_i}{k_BT}\right)^2
      \frac{k_BT}{4\lambda_i}\right]} /
    \left[1 + \exp{(x)}\right] dx
\end{equation}
\begin{equation}
  k_{Mi} = 
    \frac{\Gamma}{\hbar} \sqrt{\frac{k_BT}{4\pi\lambda_i}} \int 
    \exp{\left[-\left(x-\frac{\lambda_i - \mu_M + eE_i}{k_BT}\right)^2
      \frac{k_BT}{4\lambda_i}\right]} /
    \left[1 + \exp{(x)}\right] dx
\end{equation}
\end{subequations}
where 
\begin{equation}
  \Gamma = 2\pi \left\langle|H_{iM}|^2\right\rangle \rho
\end{equation}
is the average interfacial coupling of the site $i$ with electrode $M$, assuming the wide band approximation, $\mu_M$ $(M = L,R)$ are Fermi energies on the left ($L$) and right ($R$) electrode, respectively,
\begin{equation}
  \mu_L = eV/2,
  \qquad
  \mu_R = -eV/2
\end{equation}
distributed equally around the initial equilibrium Fermi energy $\mu_M = 0$, and $E_i$ represents the potential on the $i$-th site.


\subsubsection{Linear-chain hopping model}

In the linear model, illustrated in Fig.~1b in the main text, we assume that the four-site protein is standing on the surface and the heme cofactors connect the two electrodes like a molecular wire.
Each electrode is thus in contact with one redox site only.
The hopping model for a linear chain of $N$ redox sites connecting the left and right electrodes is described by a set of equations for the current flux $J_{ji}$ in the $i\to j$ direction,
\begin{subequations}\label{eqn_master_eq}
\begin{equation}
  J_{ji} = k_{ji} P_i (1-P_j) - k_{ij} P_j (1-P_i); \quad i,j \in \{1,\dots,4\}
\end{equation}
\begin{equation}
  J_{1L} = k_{1L} f_L (1-P_1) - k_{L1} P_1 (1-f_L)
\end{equation}
\begin{equation}
  J_{R4} = k_{R4} P_4 (1-f_R) - k_{4R} f_R (1-P_4), 
\end{equation}
\end{subequations}
where the electron transfer rates $k_{ij}$ and $k_{iM}$, $M = L,R$ are described by \eqref{eqn_k_marcus} and \eqref{eqn_k_chidsey}.
At steady state all the equations \eqref{eqn_master_eq} are equal to zero and the populations $P_i$ no longer change.
The steady-state populations can be expressed as
\begin{subequations}
\begin{equation}
  P_i = \frac{
    k_{i,i-1}P_{i-1} + k_{i,i+1}P_{i+1}}{
    k_{i-1,i} + k_{i,i+1} + P_{i-1}(k_{i,i-1}-k_{i-1,i}) + 
    P_{i+1}(k_{i,i+1}-k_{i+1,i})}
\end{equation}
\begin{equation}
  P_1 = \frac{
    k_{1L}f_L + k_{12}P_2}{
    k_{L1} + k_{21} + f_L(k_{1L}-k_{L1}) + P_2(k_{12}-k_{21})}
\end{equation}
\begin{equation}
  P_4 = \frac{
    k_{43}P_{3} + k_{4R}f_R}{
    k_{34} + k_{R4} + f_R(k_{4R}-k_{R4}) + P_{3}(k_{43}-k_{34})}
\end{equation}
\end{subequations}
These equations are coupled and therefore they are solved iteratively.
The steady--state electronic flux is then obtained from the converged site populations as
\begin{equation}
  J = J_{1L} = J_{R4}
\end{equation}
using Eq.~\ref{eqn_master_eq}.
Finally, the electric current is defined as $I = -e N_c J$ where $N_c$ is the number of protein contacts in the experimental measurements.

The asymmetry of the potential distribution is in this model included via the interfacial potential drops 
$-\alpha_LV$ and $-\alpha_RV$ between the left electrode and the first site and between the last (4th) site and the right electrode, respectively, where $\alpha_L$, $\alpha_R$ are asymmetry scaling factors.
The drop of the potential on the heme sites is assumed to be proportional to their distance from the electrodes, 
\begin{equation}
  E_i = E_i^0 + \left[\frac{1}{2} - \alpha_L - 
        (1-\alpha_L-\alpha_R)\frac{r_{1i}}{r_{1N}}\right]V, 
\end{equation}
where $E_i^0$ is the equilibrium, i.e. open-circuit, redox potential of site $i$ and $r_{ij}$ are the distances between the iron centers on the sites $i$ and $j$. 
For the fitting of the experimental $I$-$V$ curve, we consider all the equilibrium redox potentials to have the same value $E_i^0 = E^0$, and $E^0$ is treated as a fitting parameter. 
The same approximation is applied for the reorganization free energies $\lambda_{ij} = \lambda$ while the interfacial reorganization is assumed to be half of the site-site value, $\lambda_i = \lambda/2$. 
We note that experimental redox potentials and computed reorganization free energy are available for STC in aqueous solution but not for STC in vacuum, which is why we prefer to take these energies as fitting parameters.   
Values of site-site electronic couplings are taken from our previous DFT calculations on STC protein~\cite{Jiang2017}, listed in Table~\ref{tab_hopping} for convenience. 
Here we assume that the couplings are the same in solution and in vacuum because the heme position are not expected to change significantly. 
The model is therefore described by 6 free parameters: $E^0$, $\lambda$, $\alpha_L$, $\alpha_R$, $\Gamma$ and $N_c$.
The best fit to the experimental $I$-$V$ curve shown in Fig.~1a in the main text gives $E^0 = 0.01$~eV, $\lambda = 0.22$~eV, $\alpha_L = 0.025$, $\alpha_R = 0.075$, $\Gamma = 9.0$~meV, $N_c = 100$.


\subsubsection{Two-path hopping model}

In the two-path hopping model, illustrated in Fig.~1c in the main text, we assume that the protein is lying on the gold surface with the first and the last heme cofactor in contact with the right (upper) electrode and the second and third hemes are attached to the left (lower) electrode.
The model is described by the following kinetic equations,
\begin{subequations}\label{eqn_master_eq_2}
\begin{equation}
  J_{ji} = k_{ji} P_i (1-P_j) - k_{ij} P_j (1-P_i); \quad i,j \in \{1,\dots,4\}
\end{equation}
\begin{equation}
  J_{iL} = k_{iL} f_L (1-P_i) - k_{Li} P_i (1-f_L); \quad i \in \{2,3\}
\end{equation}
\begin{equation}
  J_{Ri} = k_{Ri} P_i (1-f_R) - k_{iR} f_R (1-P_i); \quad i \in \{1,4\}
\end{equation}
\end{subequations}
and the corresponding steady--state populations can be expressed as
\begin{subequations}\label{eqn_hop_p_two}
\begin{equation}
  P_1 = \frac{
    k_{1R}f_R + k_{12}P_2}{
    k_{R1} + k_{21} + f_R(k_{1R}-k_{R1}) + P_2(k_{12}-k_{21})}
\end{equation}
\begin{equation}
  P_2 = \frac{
    k_{2L}f_L + k_{21}P_1 + k_{23}P_3}{
    k_{L2} + k_{12} + k_{32} + f_L(k_{2L}-k_{L2}) + 
    P_1(k_{21}-k_{12}) + P_3(k_{23}-k_{32})}
\end{equation}
\begin{equation}
  P_3 = \frac{
    k_{3L}f_L + k_{32}P_2 + k_{34}P_4}{
    k_{L3} + k_{23} + k_{43} + f_L(k_{3L}-k_{L3}) +
    P_2(k_{32}-k_{23}) + P_4(k_{34}-k_{43})}
\end{equation}
\begin{equation}
  P_4 = \frac{
    k_{4R}f_R + k_{43}P_3}{
    k_{R4} + k_{34} + f_R(k_{4R}-k_{R4}) + P_3(k_{43}-k_{34})}.
\end{equation}
\end{subequations}
Again, the steady-state populations obtained by iterative solution of the equations \eqref{eqn_hop_p_two} define the steady--state flux
\begin{equation}
  J = J_{2L} + J_{3L} = J_{R1} + J_{R4}
\end{equation}
and the electric current is then $I = -e N_c J$.

In constrast to the linear-chain hopping model, the electrode-site distances are not considered here and the potential drop on the heme sites is governed only by the interfacial factors $\alpha_L$, $\alpha_R$.
Therefore, the actual redox potentials are 
\begin{subequations}
\begin{equation}
  \textstyle
  E_1 = E_1^0 - \left[\frac{1}{2} - \alpha_R\right]V
\end{equation}
\begin{equation}
  \textstyle
  E_2 = E_2^0 + \left[\frac{1}{2} - \alpha_L\right]V
\end{equation}
\begin{equation}
  \textstyle
  E_3 = E_3^0 + \left[\frac{1}{2} - \alpha_L\right]V
\end{equation}
\begin{equation}
  \textstyle
  E_4 = E_4^0 - \left[\frac{1}{2} - \alpha_R\right]V
\end{equation}
\end{subequations}
As in the linear-chain model, there are 6 free fitting parameters in the two-path hopping model: $E^0$, $\lambda$, $\alpha_L$, $\alpha_R$, $\Gamma$ and $N_c$. 
The best fit to the experimental $I$-$V$ curve shown in Fig.~1a in the main text gives $E^0 = 0.0$~eV, $\lambda = 0.30$~eV, $\alpha_L = 0.125$, $\alpha_R = 0.443$, $\Gamma = 500$~meV, $N_c = 25$.


\subsection{Coherent tunneling models}\label{sec_coh_tunnel}

The coherent tunnelling current through a potential barrier separating two conductive electrodes can be written as 
\begin{equation}\label{eqn_tunnel_i}
  I(V) = \frac{e}{\pi\hbar} 
    \int_{-\infty}^{\infty} T(E) \left[ f_L(E) - f_R(E) \right] dE
\end{equation}
where $T(E)$ is the transmission function, that is the probability that the electron of energy $E$ tunnels through a dielectric medium (barrier) between the electrodes.~\cite{Nitzan2006}
The probability is integrated over a Fermi window, defined as the difference of the Fermi--Dirac distributions
\begin{equation}
  f(E) = \frac{1}{\exp{\left(\frac{E-\mu}{k_BT}\right)}+1}
\end{equation}
of the left ($f_L$) and right ($f_R$) electrodes. Obviously, the current is non-zero only in a non-equilibrium state, that is when the Fermi levels on the left ($\mu_L$) and right ($\mu_R$) electrodes are not equal.


\subsubsection{Simmons model}

The model of Simmons~\cite{Simmons1963} (Fig.~1d in the main text) describes electron tunnelling through a potential barrier of arbitrary shape representing a thin insulating film between two conductive electrodes. 
Assuming the average barrier height $\phi$ above the Fermi level $\mu_L$ of the negatively-charged left electrode and the potential drop occurring 
at the right electrode, the current can be described as
\begin{equation}\label{eqn_simmons_i}
  I(V) = 
    \frac{e^2}{2\pi h} \left[
      \left(\phi - \alpha V\right) 
         e^{-K\sqrt{\phi - \alpha V}} -
      \left(\phi + \left(1-\alpha\right)V\right) 
         e^{-K\sqrt{\phi + \left(1-\alpha\right)V}}
    \right]
\end{equation}
where
\begin{equation}
  K = 4\pi L\left(2me\right)^{1/2}/h
\end{equation}
is determined by tunnelling length $L$.
The model involves also the so-called symmetry factor $\alpha$ allowing description of arbitrary distribution of the potential drop on the electrodes.~\cite{Zhao2004}
Here, we use the formula for the fitting of the experimental $I$-$V$ curves.
As the experimental measurements proceed on certain, a priori unknown, number of protein contacts $N_c$, we introduce this parameter into \eqref{eqn_simmons_i} as a scaling factor. 
The modelling therefore proceeds with four free parameters: $\phi$, $L$, $\alpha$ and $N_c$.
Optimized values of the parameters describing the curve shown in Fig.~1a in the main text are $\phi = 1.13$~eV, $L = 12.2$~\AA, $\alpha = 0.42$ and $N_c = 20347$.
The obtained parameter values are essentially the same as in previous work of Garg \emph{et al.}~\cite{Garg2018} (note that the Eq.~3 in Ref.~\citenum{Garg2018} might be misleading, the $(2me)^{1/2}$ factor in exponential factor $K$ is in the numerator not in the denominator) where the current density characterized by effective surface area was modelled instead of current scaled by effective number of contacts as here.
The large number of contacts $N_c$ in the current fit corresponds to 10-times larger effective area than the surface of the nanowire used in Ref.~\citenum{Garg2018}.
Although the model reproduces the I-V curve exceptionally well, these unrealistic values make interpretation of the Simmons-model parameters difficults and therefore we applied more sophisticated Landauer model.


\subsubsection{Landauer model}

In the Landauer formalism~\cite{Carey2017, Valianti2019}, the transmission function in \eqref{eqn_tunnel_i} is approximated by a 
Lo\-ren\-tzian function
\begin{equation}\label{eqn_landauer_t}
  T(E) = \frac{\Gamma_L\Gamma_R}{
    [E - \epsilon_0]^2 + \Gamma^2},
  \qquad
  \Gamma = (\Gamma_L+\Gamma_R)/2
\end{equation}
representing the conduction channel, that is a molecular energy level $\epsilon_0$ mediating the tunnelling current.
$\Gamma_L$ and $\Gamma_R$ are spectral densities (interfacial protein/electrode couplings exhibited by interfacial state broadening) determining the shape of the transmission function (Fig.~1e in the main text).

As in the Simmons model, a symmetry factor $\alpha$ is introduced to control the potential drop on the electrodes via the positions of their Fermi levels
\begin{equation}
  \mu_L = \alpha eV/2, 
  \qquad
  \mu_R = -(1-\alpha) eV/2
\end{equation}
The integral \eqref{eqn_tunnel_i} with transmission function \eqref{eqn_landauer_t} is solved in the zero-temperature limit, where the Fermi--Dirac distribution converges to a Heaviside step function $f(E) \to \vartheta(E-\mu)$ and $df/d\mu \to \delta(E-\mu)$. 
In this limit the electric conductance can be written in analytic form,
\begin{equation}\label{eqn_landauer_g}
  \begin{split}
  g(V) 
    &= \frac{dI}{dV} 
     = \frac{e}{\pi\hbar} \left[
       \int T(E) \frac{df}{d\mu_L}\frac{d\mu_L}{dV} dE -
       \int T(E) \frac{df}{d\mu_R}\frac{d\mu_R}{dV} dE \right] = \\
    &= G_0 \Gamma_L\Gamma_R \left[
       \frac{\alpha}{(\alpha eV-\epsilon_0)^2 + \Gamma^2} -
       \frac{\alpha-1}{((\alpha-1)eV-\epsilon_0)^2 + \Gamma^2}
       \right],
  \end{split}
\end{equation}
where we introduced the quantum conductance $G_0 = e^2/\pi\hbar$. 
The tunneling current is then obtained by integration of \eqref{eqn_landauer_g}, giving
\begin{equation} \label{eqn_fit}
  I(V) = 
    N_c  \frac{2 G_0}{e} \frac{\Gamma_L\Gamma_R}{\Gamma_L + \Gamma_R} \left[
    \arctan{\frac{\alpha eV - \epsilon_0}{(\Gamma_L + \Gamma_R)/2}} -
    \arctan{\frac{(\alpha-1)eV - \epsilon_0}{(\Gamma_L + \Gamma_R)/2}}
    \right], 
\end{equation}
where we have introduced a multiplicative factor $N_c$ that equals the number of protein contacts in the junction. 
For the fitting of the experimental $I$-$V$ curve to \eqref{eqn_fit}, we assume the same coupling to the left and right electrode 
($\Gamma_L = \Gamma_R \equiv \Gamma$). 
Hence, there are four free fitting parameters: $\epsilon_0$, $\Gamma$, $\alpha$ and $N_c$.
The best fitting parameters for the experimental $I$-$V$ curve shown in Fig.~1a in the main text take values 
$\epsilon_0 = 0.56$~eV, $\Gamma = 6.21$~meV, $\alpha = 0.41$, $N_c = 68$.


\section{Atomistic structure of the Au-STC-Au junction}


\subsection{Initial structure of STC}

The starting point is the X-ray crystallographic structure of the all-oxidized small tetraheme cytochrome (STC) from \emph{Shewanella oneidensis}, at a resolution of 0.97 \AA~\cite{Leys2002} (PBD ID 1M1Q). 
The STC protein was described by the CHARMM27 force field~\cite{MacKerell1998, MacKerell2004} supplemented with the parameters for heme cofactors, axial histidines and cysteine linkages from our previous studies of multi-heme proteins~\cite{Breuer2012, Breuer2014, Jiang2017}. 
Crystallographic water molecules were described by TIP3P model~\cite{Jorgensen1983}.
In accord with the experimental setup, serine SER87 near the C-terminus was mutated to cysteine for chemisorption of the protein to the gold surface at a later stage of the structure preparation.
Since experimental measurements on the junction were carried out in vacuum, amino acids at the protein surface were kept neutral, except Lys-20 -- Asp-21 and Asp-81 -- Arg-83 pairs which form salt bridges.
The N- and C-termini of the amino-acid chain were terminated by acetyl (ACE) and amine group (NH2), respectively.
All heme cofactors were oxidized in accord with the experimental oxidation state of Fe in the as-measured junctions, i.e. they bear Fe$^{3+}$ cations coordinated to the porphyrin-ring and two axial histidines. 
One of the two propionate side chains on each heme was protonated (-(CH$_2$)$_2$-COOH) while the other was ionized (-(CH$_2$)$_2$-COO$^-$) to keep the net charge of each cofactor equal to zero.
In this way, the whole protein was charge-neutral and no counter ions were added.


\subsection{MD simulation of STC in vacuum}

The crystal structure of STC contains 272 crystallographic water molecules. 
Under experimental conditions (high vacuum, 10$^{-5}$ bar)\cite{Garg2018}, the protein is not solvated, yet some water molecules are expected to be present within the protein or in protein surface pockets due to strong hydrogen bonding with protein residues. 
To determine the position and stability of these structural water molecules, we simulated water evaporation while tracking the structural changes.

\begin{figure}[t]
  \centering
  \includegraphics[scale=1.0]{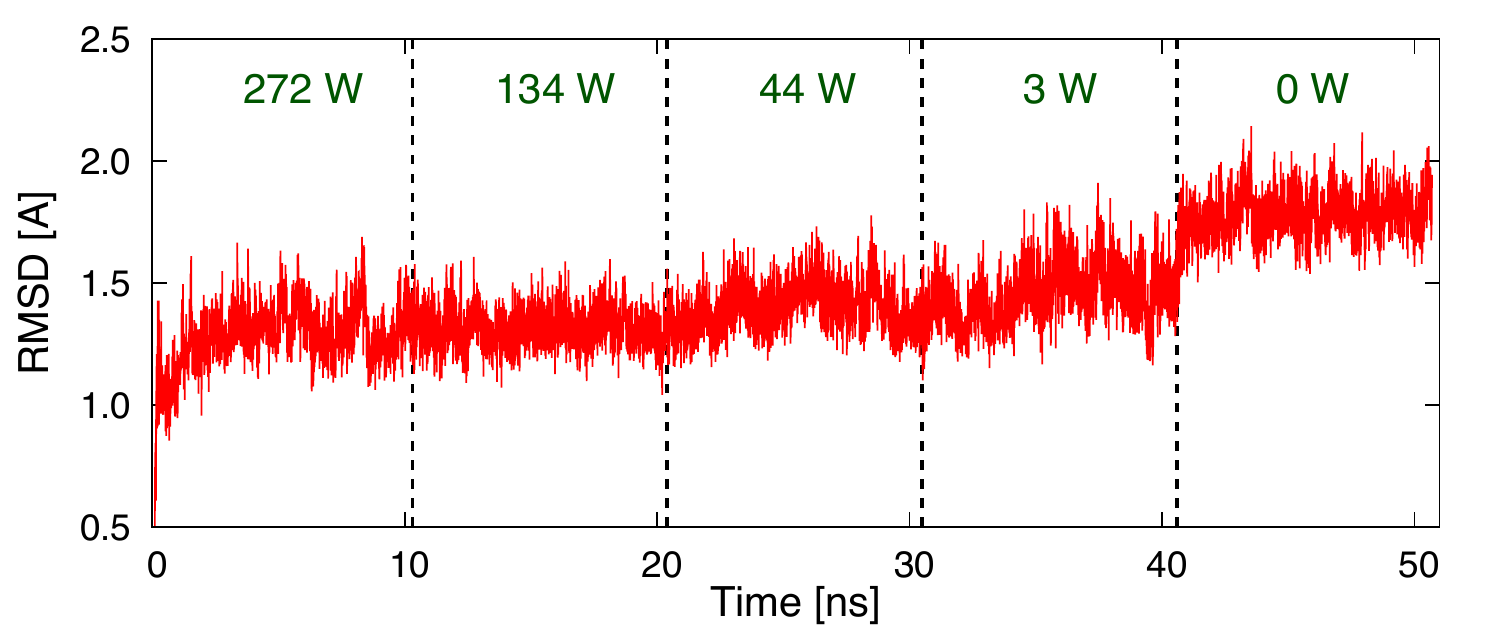}
  \caption{
RMSD of STC protein backbone atoms with respect to crystal structure (PBD ID: 1M1Q) along MD trajectories in vacuum. 
Crystallographic water molecules evaporate from the structure and are sequentially removed from the simulation as indicated.}
  \label{fig_vac_rmsd}
\end{figure}

The protein and all crystallographic water molecules were placed into the center of a $14.6\!\times\!15.3\!\times\!15.5$ nm$^3$ simulation box applying periodic boundary conditions (PBC).
The positions of hydrogen atoms and water molecules were first relaxed by a 5 ps molecular dynamics (MD) run with time step of 0.25 fs at a temperature of 150 K controlled by velocity-rescaling algorithm, while all the protein heavy atoms were constrained in their initial coordinates. 
Then, a series of four 50 ps MD runs was performed with a weak-coupling Berendsen thermostat~\cite{Berendsen1984} ($\tau$ = 1 ps, $T$ = 300 K) applied to water molecules and protein hydrogen atoms. 
The positions of the protein heavy atoms were constrained in the first run and restrained in the following three runs by applying a harmonic potential with force constants 1000, 500 and 50 kJ/mol/nm$^2$. 
The equilibration stage was finished with two MD simulations with free, non-restrained systems kept at a temperature of 300 K. 
In the first 50 ps run, the Berendsen thermostat ($\tau$ = 1 ps) was applied on 
all atoms, and in the second 100 ps run a Nose-Hoover thermostat~\cite{Nose1984, Hoover1985} ($\tau$ = 1 ps) was used. 
Finally, a production run of 10 ns was carried out with a Nose-Hoover thermostat during which the root-mean-square deviation (RMSD) of the protein backbone with respect to the crystal structure was tracked 
(shown in Fig.~\ref{fig_vac_rmsd}).

Water molecules that evaporated from the protein surface during the production run were removed from the structure and the system was again equilibrated by 100 ps MD run followed by the 10 ns production run with the same setup as described above. 
This procedure was repeated four times reducing sequentially the number of water molecules from initially 272 to 134, 44, 3 and finally to the system without any water. 
The RMSD as a function of MD time shown in Fig.~\ref{fig_vac_rmsd} indicates that the protein remains folded in a stable configuration as long as three structural water molecules are retained in the system. 
When these are removed the more flexible surface loops become distorted, which is manifested by an abrupt increase of the RMSD. 
Therefore we keep these three structural water molecules in our STC model.

All MD simulations reported in this work used a smooth particle mesh Ewald (SPME) method for evaluation of the electrostatic interaction under PBC~\cite{Essmann1995}. 
The PME real-space cutoff was set to 12 \AA\ while the Fourier-space grid was constructed with a spacing of 1.5 \AA\ and interpolated by cubic polynoms.
Short-range Lennard-Jones (LJ) interactions were smoothly attenuated to zero between 11 \AA\ and 12 \AA.
All MD simulations were performed with Gromacs 5.1.4~\cite{Berendsen1995, vanderSpoel2005, Abraham2015}.


\subsection{Adsorption of STC on Au(111)}\label{sec_surf_ads}

The protein structure including the 3 structural water molecules was relaxed in vacuum by additional 10 ns long MD run at 300 K and placed in a $99.620\!\times\!98.961\!\times\!111.962$ \AA$^3$ simulation cell as detailed below. 
The bottom of the cell contained a 6-monolayer (ML) thick Au(111) surface slab (7956 Au atoms). 
The lattice constant of 4.144 \AA\ was used to construct the gold slab which corresponds to a Au--Au distance of 2.93 \AA. 
Due to the periodic boundary conditions applied in all 3 directions there was a 100 \AA\ separation distance between the gold surface and its image.  
The gold surface was described by the GolP-CHARMM force field,~\cite{Iori2008, Iori2009, Wright2013} which is polarizable and can capture the image-charge interaction between gold and adsorbed molecular species. 
Although the GolP-CHARMM force field was parametrized for interactions with aqueous species, we have shown in previous work that this force field also predicts 
adsorption structures and energies of amino acids in vacuum in good agreement with accurate vdW-DF reference data~\cite{Hammer1999, Dion2004} (mean relative unsigned error in adsorption energies = 11.3\%)~\cite{Futera2019}.

The protein was placed above the gold surface in a way that the shortest contact distance measured from the surface of the protein to the top-most gold layer was 10~\AA.
To generate different initial orientations of the protein in a systematic way, we defined an axis pointing from the protein center of mass to the sulfur atom of Cys-87 near the C-terminus and aligned it with the gold-surface normal defined to be the $z$-direction.
The protein orientation is determined by three Euler angles which we define as $\alpha$ (rotation around the $z$ axis), $\beta$ (rotation around the $x$ axis) and $\gamma$ (rotation around the protein axis), applied in this order.
Since in the experimental setup the STC protein is adsorbed on the surface by Cys-87 we scanned only the half-sphere of initial orientations with the protein axis pointing to the surface: $\alpha \in \{0, 15, 30, 45\}$ deg, $\beta \in \{90, 108, 126, 144, 162, 180\}$ deg and $\gamma \in \{0, 60, 120, 180, 240, 300\}$ deg.

\begin{figure}[tbh]
  \centering
  \includegraphics[scale=1.0]{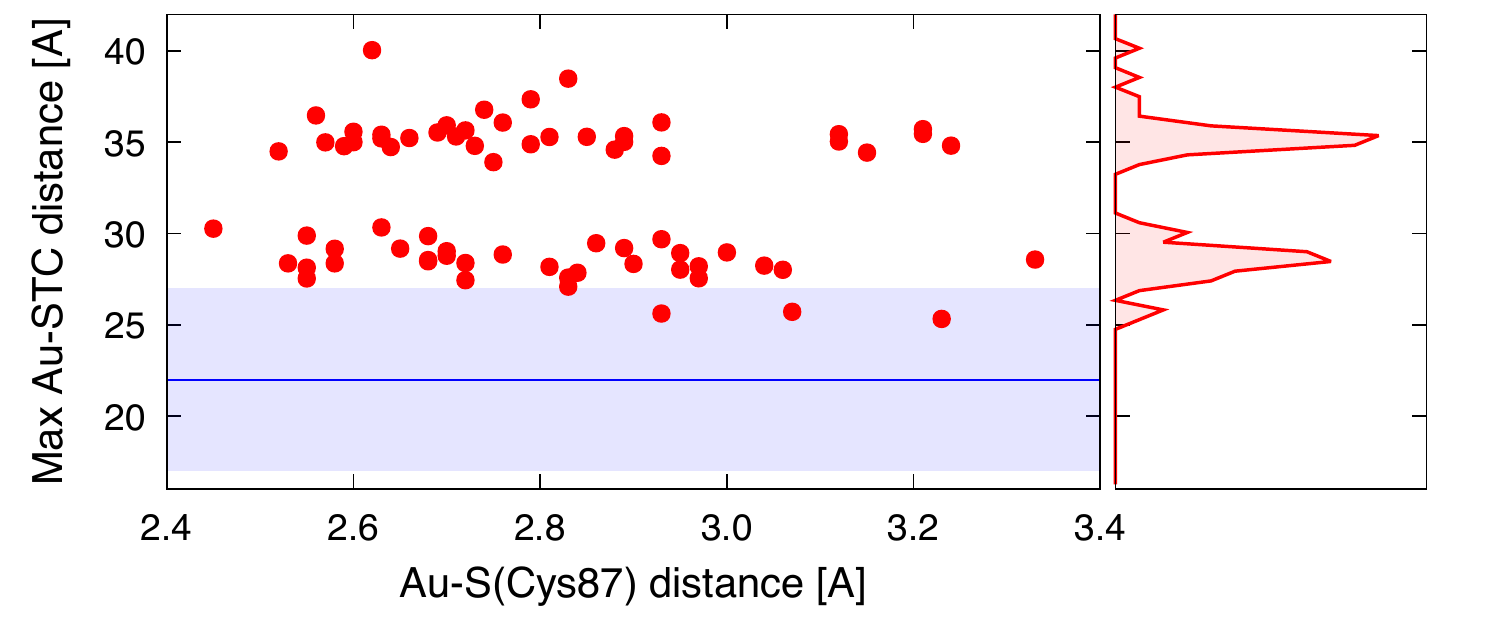}
  \caption{
Correlation plot of the most distant STC atoms from the gold-surface plane with the Au-S(Cys87) distance (red dots). 
Experimental mean value of the protein monolayer width (solid blue line) and error bar (ligth blue band) are shown for comparison. 
A histogram of the plotted Au-STC distances is shown in the right-hand-side panel.}
  \label{fig_ads_stat}
\end{figure}

Each of the 144 initial structures was relaxed in a 0.5 ps MD run (timestep $\Delta t$ = 0.1 fs) at a temperature of 150 K imposed by velocity rescaling, followed by a 5 ps MD run ($\Delta t$ = 0.5 fs) at $T$ = 300~K (Berendsen thermostat, $\tau$ = 0.5 ps) and a subsequent 10 ps MD run ($\Delta t$ = 1 fs) at  $T$ = 300~K (Nose-Hoover thermostat, $\tau$ = 0.5 ps).
A production run of 10 ns MD with the latter setup followed the equilibration. 
The positions of all the surface gold atoms were frozen in all MD runs while the ensemble of the gold-slab rigid-dipoles was propagated at $T$ = 300~K, regardless of the temperature of the rest of the system, as is customary for GolP-CHARMM simulations~\cite{Iori2008, Iori2009, Wright2013}.
All simulated samples spontaneously adsorbed on the gold surface during the production MD simulations.
71 out of the 144 structures (49.3\%) adsorbed with the sulfur atom of Cys-87 on the surface (see Fig.~\ref{fig_ads_stat}). 
For further modelling of the junction, we selected a structure that was ``lying" on the gold surface and that gave a layer thickness in good agreement with experiment. 
For a discussion of the adsorption structures generated we refer to the main text.   


\subsection{Model structure of Au-STC-Au junction}

The selected physisorbed STC structure was immobilized on the gold surface by modelling the interaction Au-S(Cys-87) by a harmonic bond with force field parameters taken from Ref.~\citenum{Futera2019}. 
The height of the simulation box was decreased to 27.3 \AA\ so that the distance between the protein and the closest periodic image of the gold surface slab in 
$z$-direction was 2.0 \AA.
Like in the previous classical simulations, the periodic boundary conditions were applied in all 3 directions here and only one explicit gold slab was present in the simulation box.
The electrode separation distance was then further decreased to 22.0 \AA\ in steps of 0.25 \AA. 
In each step the protein coordinates were rescaled assuming isotropic compression, the structure was equilibrated using the procedure described in Section~\ref{sec_surf_ads}, and relaxed by 1 ns MD simulation. 
To establish the equilibrium electrode separation, we evaluated the local pressure acting on the protein along the surface-normal direction by integrating the average local stress tensor in the protein region~\cite{Thompson2009, Ollila2009}. 
We obtained a value of 26.8~\AA\ for the electrode separation that gives a local pressure in the protein that is close to zero. 
The structure at this distance was relaxed by 1 ns long MD and the last snapshot of the trajectory was used for the DFT calculations described below.


\section{DFT calculations on the Au-STC-Au junction}\label{sec_dft_model}


\subsection{Kohn-Sham DFT calculations}

\begin{figure}[tbh]
  \centering
  \begin{picture}(431,216)
    \put(  0,  0){\includegraphics[width=420pt]{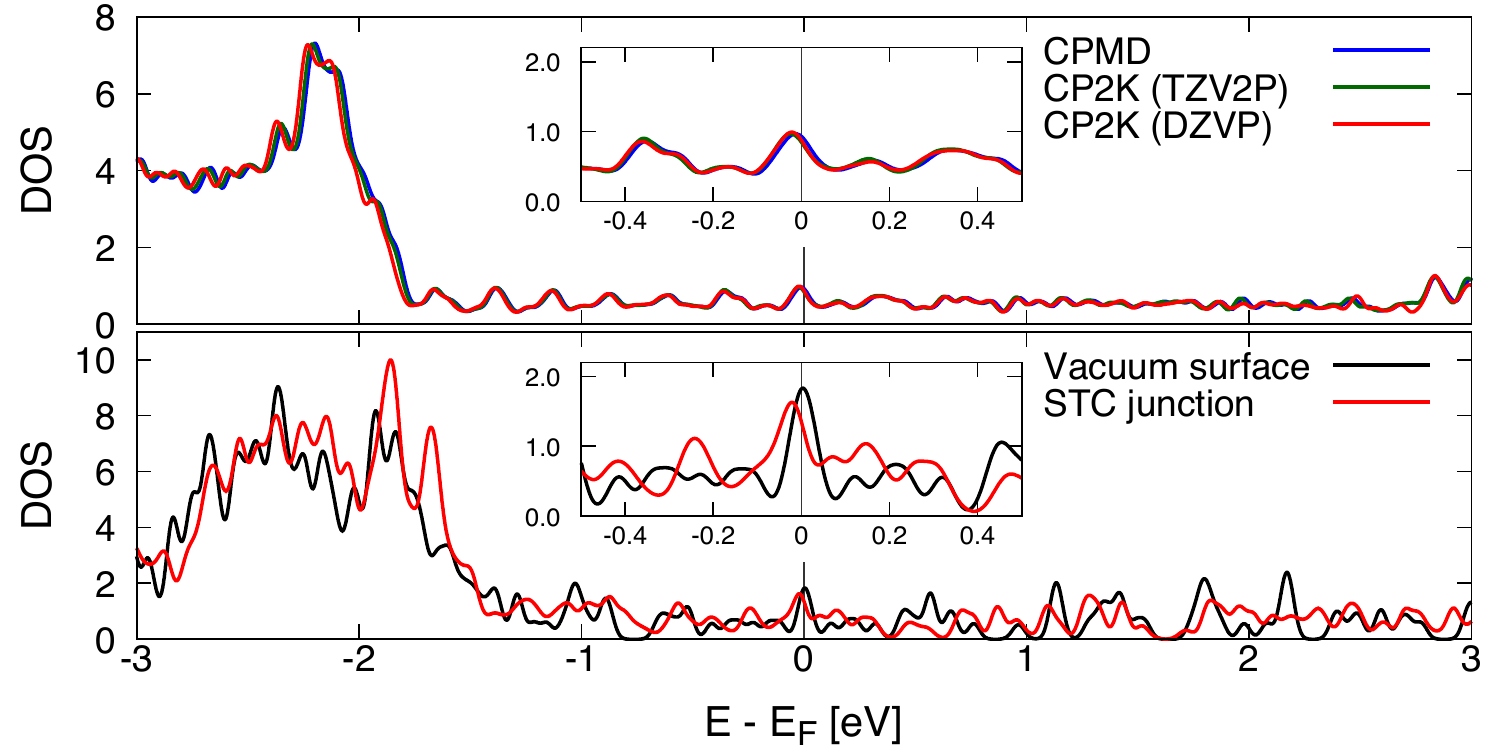}}
    \put(415,120){(a)}
    \put(415, 10){(b)}
  \end{picture}
  \caption{
Density of gold states (DOS) calculated (a) on bulk gold structure using plane-wave CPMD code and the CP2K code with localized double-$\zeta$ (DZVP) and triple-$\zeta$ (TVZV2P) basis sets.
25 K-points in each dimension were used to sample the Brillouin zone in reciprocal space.
In (b), gold DOS obtained from Au-STC-Au junction (two 2-ML slabs, CP2K with DZVP, $\Gamma$ point only) is compared with 2-ML slab calculation in vacuum (CPMD, $14\!\times\!8\!\times\!3$ K-point mesh).
Energy scale is aligned to Fermi level $E_F$ and the region close to $E_F$ is zoom in the insets for clarity.}
  \label{fig_gold_dos}
\end{figure}

For the density-functional-theory (DFT) calculations, the area of the surface on which the protein is adsorbed was reduced to $49.810\!\times\!50.749$ \AA$^2$, which corresponds to $\sim$10 \AA\ distance between the protein 
and its nearest periodic image on either side. 
The number of gold monolayers was reduced to 2 for each electrode as a compromise between accuracy and computational cost of the DFT calculations. 
Although larger number of gold monolayers cannot be explicitly included in the model because of its prohibitive computational demandness, we have verified that the density of gold states is well represented in 2-ML model due to the large surface area (see Fig.~\ref{fig_gold_dos}).
In contrast to classical simulations, the DFT model is periodic only in the plane of the gold surface ($x,y$-directions) but non-periodic in the surface-normal direction ($z$-direction). 
Therefore two explicit Au(111) slabs were present in the supercell, representing left and right electrode of the protein junction. 
Each side of the supercell in the $z$ direction was padded by 9.2 \AA\ vacuum region leading to supercell $z$-dimension 49.954 \AA. 
The total number of protein (including 3 water molecules) and gold atoms for which DFT calculations were performed are 1569 and 1360, respectively.

\begin{figure}[tbh]
  \centering
  \includegraphics[scale=1.0]{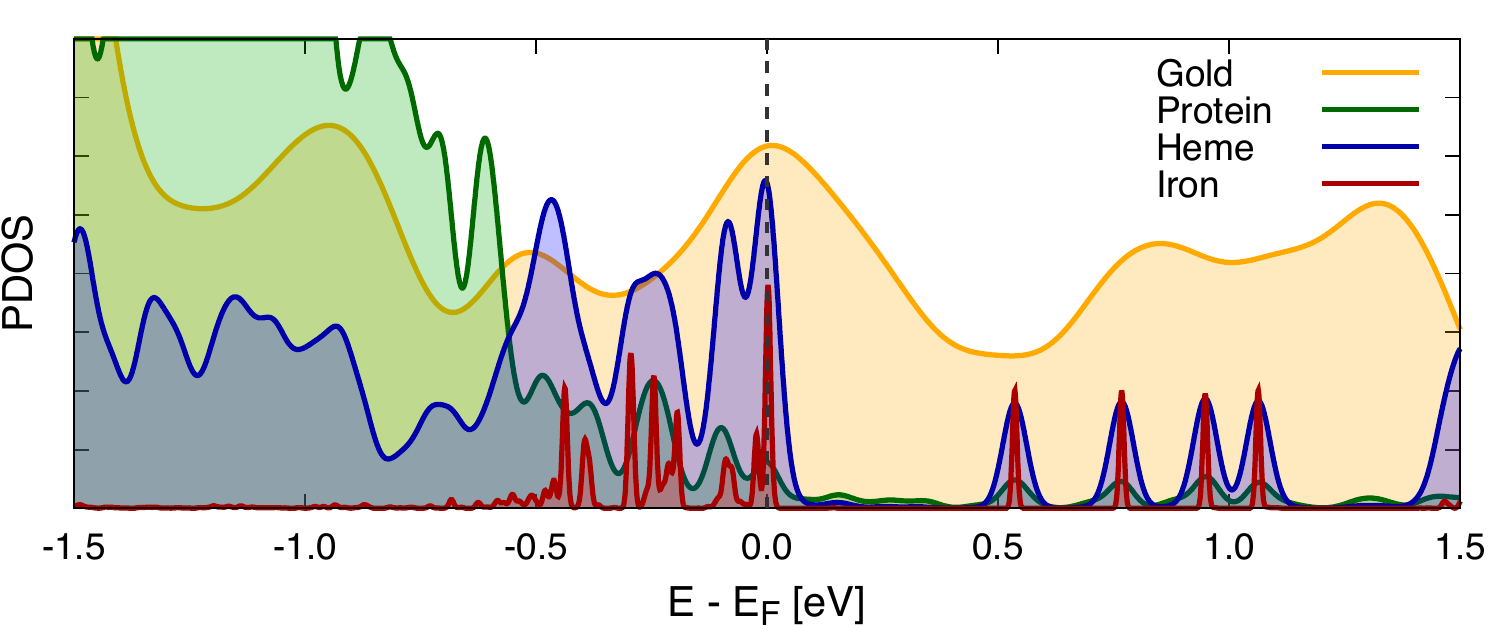}
  \caption{
Projected density of states (PDOS) of STC-gold junction at PBE level of theory.
Contributions from Fe atoms (Iron), the porphyrine rings and axial histidines coordinates to Fe (Heme), all protein amino acids except the axial histidines (Protein) and gold are shown.}
  \label{fig_pdos}
\end{figure}

DFT calculations were carried out using the CP2K~\cite{Hutter2014} software package and the PBE functional~\cite{Perdew1996}.
While the core electrons were replaced by Goedecker-Teter-Hutter (GTH) pseudopotentials~\cite{Goedecker1996}, a double-$\zeta$ basis set with polarization functions from the CP2K database (DZVP-MOLOPT-SR) was used for expansion of the orbitals. 
The total number of electrons was 19547 (two gold electrodes: 14960 and protein: 4587 electrons) and the total number of basis functions was 48569.
Fermi smearing with an electronic temperature of 300 K was applied to maintain fractional occupation of metallic states near the Fermi level. 
A cutoff of 500 Ry was used to define the finest multigrid used for real-space integration. 
The wavefunction was optimized using Broyden's mixing scheme~\cite{Broyden1965} to an accuracy of 10$^{-6}$ a.u. 
The projected density of Kohn-Sham states of the Au-STC-Au junction is shown Figure~\ref{fig_pdos}.  


\subsection{Orbital localization, orbital energy and protein-electrode electronic coupling} 

We employed the projection-operator diabatization (POD) method~\cite{Futera2017}, implemented in CP2K software package by our group, to localize electronic states on the system fragments (two gold electrodes and STC protein) and evaluate electronic coupling elements.
The POD method constructs the localized states from the Kohn-Sham (KS) adiabatic states
\begin{equation}
  \hat{H} |\psi_i\rangle = 
    \epsilon_i |\psi_i\rangle,
\end{equation}\label{eq_ks}
which are obtained by standard SCF calculations.
The KS Hamiltonian is transformed to orthonormalized basis set of atom-center localized functions $\{\phi_i\}$ and rearranged to the following block structure
\begin{equation}
  \tilde{H} = \left[
    \begin{array}{ccc}
      \tilde{H}_{LL} & \tilde{H}_{LP} & \tilde{H}_{LR} \\
      \tilde{H}_{PL} & \tilde{H}_{PP} & \tilde{H}_{PR} \\
      \tilde{H}_{RL} & \tilde{H}_{RP} & \tilde{H}_{RR} \\
    \end{array}
    \right]
\end{equation}\label{eq_h_blocks}
where labels $L$, $R$ and $P$ and stand for the left electrode, right electrode and the protein, respectively.
The desired localized states are obtained by diagonalization of the diagonal blocks, $\tilde{H}_{\alpha\alpha}$, $\alpha\in\{L,R,P\}$, while the off-diagonal blocks are transformed correspondigly:
\begin{gather}
  \bar{H}_{\alpha\alpha} = 
    U_\alpha^\dagger \cdot \tilde{H}_{\alpha\alpha} \cdot U_\alpha, \\
  \bar{H}_{\alpha\beta} = 
    U_\alpha^\dagger \cdot \tilde{H}_{\alpha\beta} \cdot U_\beta.
\end{gather}\label{eq_block_diag}
After this unitary transformation the system Hamiltonian has the following form
\begin{equation}
  \bar{H} = \left[
  \begin{array}{ccccccccc}
    \epsilon_{L,1} & \dots & 0 \\
    \vdots & \ddots & \vdots & & \bar{H}_{LP} & & & \bar{H}_{LR} \\
    0 & \dots & \epsilon_{L,N_L} \\
    & & & \epsilon_{P,1} & \dots & 0 \\
    & \bar{H}_{PL} & & \vdots & \ddots & \vdots & & \bar{H}_{PR} \\
    & & & 0 & \dots & \epsilon_{P,N_P} \\
    & & & & & & \epsilon_{R,1} & \dots & 0 \\
    & \bar{H}_{RL} & & & \bar{H}_{RP} & & \vdots & \ddots & \vdots \\
    & & & & & & 0 & \dots & \epsilon_{R,N_R}
  \end{array}
  \right]\label{eq_h_pod}
\end{equation}
On the main diagonal, there are $N_L$, $N_P$, $N_R$ localized-state, one-electron energies of the left-electrode, protein and the right-electrode while the off-diagonal blocks $\bar{H}_{\alpha\beta}$ contain electronic coupling elements between the corresponding states.
In our case of Au-STC-Au junction, there are $N_L = N_R = 17000$ metal states and $N_P = 14569$ protein states.

Spectral density functions characterizing interfacial metal / protein coupling are constructed from the POD electronic coupling elements of the sub-diagonal blocks in Eq.~\ref{eq_h_pod}.
\begin{equation}
  \Gamma_j^{(M)}(E) = 2\pi\sum_q^{N_M}
    \left|[\hat{H}_{MP}]_{qj}\right|^2
    \delta(E-\epsilon_{M,q})
\end{equation}
where $M\in\{L,R\}$.
Transmission function for the molecular junction has in Breit-Wigner approximation form of the sum of Lorentzian peaks characterising the delocalizaton and coupling of each protein state $j$ to the both electrodes~\cite{Carey2017, Papp2019}
\begin{equation}
  T_j(E) = 
    \frac{\Gamma_j^{(L)}\Gamma_j^{(R)}}
    {(E-\epsilon_{P,j})^2 + [\Gamma_j(E)]^2},
  \qquad
  \Gamma_j(E) = 
    \frac{1}{2}\left[
    \Gamma_j^{(L)} + \Gamma_j^{(R)}\right]
\end{equation}
The transmission functions are then used in Landauer formalism for evaluation of the tunnelling current, as it is described in the main text and in the Section~\ref{sec_coh_tunnel}.

\hl{Note, that we calculate the transmission function in the whole energy range given by the bias-dependent Fermi window and evaluate its integral in this range, in contrast to zero-temperature zero-bias formula for conductivity evaluated from the transmission on the Fermi level only, which is not used in this work.
Wide band approximation is not applied here, all state-dependent coupling elements are taken explicitly from the POD calculation.
The Breit-Wigner approximation of the transmission function used here assumes that the full all-to-all transmission matrix is diagonal and the molecular electronic states do not interact with each other. 
The all-to-all transmission matrix currently cannot be constructed for this large system because of the prohibitively large computation demand. 
However, we believe that disregarding of quantum interference in the big systems like protein junctions is a reasonable assumption, which we justify by a good agreement of the calculated $I$-$V$ curve with the experimental data.}


\subsection{Electronic band alignment}

DFT(GGA) calculations of molecule/metal interfaces suffer from two well known problems. 
First, the energy gap between the highest occupied molecular orbital (HOMO) and the lowest unoccupied molecular orbital (LUMO) of the molecule in the gas phase is underestimated due to the electron self-interaction error. 
Second, when the molecule interacts with the metallic surface the levels renormalize due to non-local image charge interactions leading to a reduction in the energy gap. 
Image-charge interactions are not properly accounted for in DFT(GGA) because of the asymptotically incorrect exchange correlation potential. 
Unfortunately the two opposing trends in the shift of HOMO and LUMO do not cancel and lead to an incorrect alignment of the molecular states with respective to the metallic states, i.e. the Fermi level of the electrode.     

Here we use the DFT+$\Sigma$~\cite{Neaton2006, Quek2007, Egger2015} approach to correct the alignment of the molecular electronic energy levels with respect to the levels of the gold electrodes. 
In general, each molecular orbital energy $\epsilon^\alpha$, here obtained 
from POD following a KS-DFT calculation, deviates from the accurate reference value $\epsilon^\alpha_0$ by an energy shift $\Sigma_\alpha$. In DFT+$\Sigma$ one writes
\begin{eqnarray}
  \epsilon_0^\alpha & =  & \epsilon^\alpha + \Sigma^\alpha \\
  \Sigma^\alpha & = &  \Sigma_0^\alpha + \Sigma_{pol}^\alpha, 
\end{eqnarray}
where $\Sigma^\alpha$ is comprised of two parts, a correction for the self-interaction error, $\Sigma_{0}^{\alpha}$, and a correction for the missing level renormalization at the interface, $ \Sigma_{pol}^{\alpha}$. 
Here we calculate the correction only for the HOMO and LUMO ($\alpha\!=\!$ HOMO, LUMO) and use it to shift all occupied and all unoccupied levels uniformly by the same amount. 
Unfortunately, the calculation of the $\Sigma^\alpha$ correction from the electronic properties of the full protein is computationally prohibitive as it requires the use of functionals containing exact exchange. 
Yet, as one can see in Fig.~\ref{fig_pdos}, the top of the protein valence band and the bottom of the conduction band are dominated by Fe-heme electronic states. 
Hence, the calculation of the $\Sigma^\alpha$ correction from the electronic properties of the bis-histidine coordinated Fe-heme cofactors rather than the full protein should give a good approximation, as we detail in the following. 


\subsubsection{HOMO-LUMO energy gap correction}

Reference electronic states for gas phase bandgap correction were calculated using the optimally tuned range-separated hybrid functional (OT-RSH)~\cite{Baer2010, Stein2010, Kronik2012}. 
In this approach, the Coulomb operator is split into short-range and long-range parts~\cite{Yanai2004}
\begin{equation}
  \frac{1}{|\mathbf{r}-\mathbf{r}'|} =
    \frac{\alpha + \beta\mathrm{erf}{(\omega|\mathbf{r}-\mathbf{r}'|)}}{
      |\mathbf{r}-\mathbf{r}'|} + 
    \frac{1 - [\alpha + \beta\mathrm{erf}{(\omega|\mathbf{r}-\mathbf{r}'|)}]}{
      |\mathbf{r}-\mathbf{r}'|}
\end{equation}
where the parameter $\alpha$ determines the fraction of Hartree-Fock exchange (HFX) in the short-range region while the fraction of HFX in the long-range region is $\alpha + \beta$. 
A smooth transition between these two regions is maintained by the error function. 
Its steepness is controlled by range-separated parameter $\omega$. 
At range-separation distance $\rho = 1/\omega$ about 84\% of the short-range fraction of HFX is replaced by GGA exchange and the HFX is completely replaced by GGA at interaction distances greater than 2$\rho$. 
In the OT-RSH approach the range separation parameter $\omega$ is optimized to enforce the exact conditions IP = $-\epsilon_N^\mathrm{HOMO}$, EA = $-\epsilon_{N+1}^\mathrm{HOMO}$, where IP = $E(N-1)-E(N)$, EA = $E(N)-E(N+1)$ are the vertical ionization potential and electron affinity, and $E(N)$ is the total electronic energy of system with $N$ electrons. 
In practice, the optimization is done by minimizing the following functional
\begin{equation}
  J(\omega) = 
    \left[\epsilon^\mathrm{HOMO}_{N,\omega} + 
      \mathrm{IP}_{N,\omega}\right]^2 +
    \left[\epsilon^\mathrm{HOMO}_{N+1,\omega} + 
      \mathrm{IP}_{N+1,\omega}\right]^2 
  \label{eq_otrsh_j}
\end{equation}
Here, PBE\cite{Perdew1996} is used for the GGA part. 
The $\omega$ parameter is optimized applying 20\% HFX in the short-range region ($\alpha = 0.2$) and 100\% HFX in the long-range region ($\beta = 1-\alpha = 0.8$) which ensures correct asymptotic behavior.

\begin{figure}[tbh]
  \centering
  \includegraphics[scale=1.0]{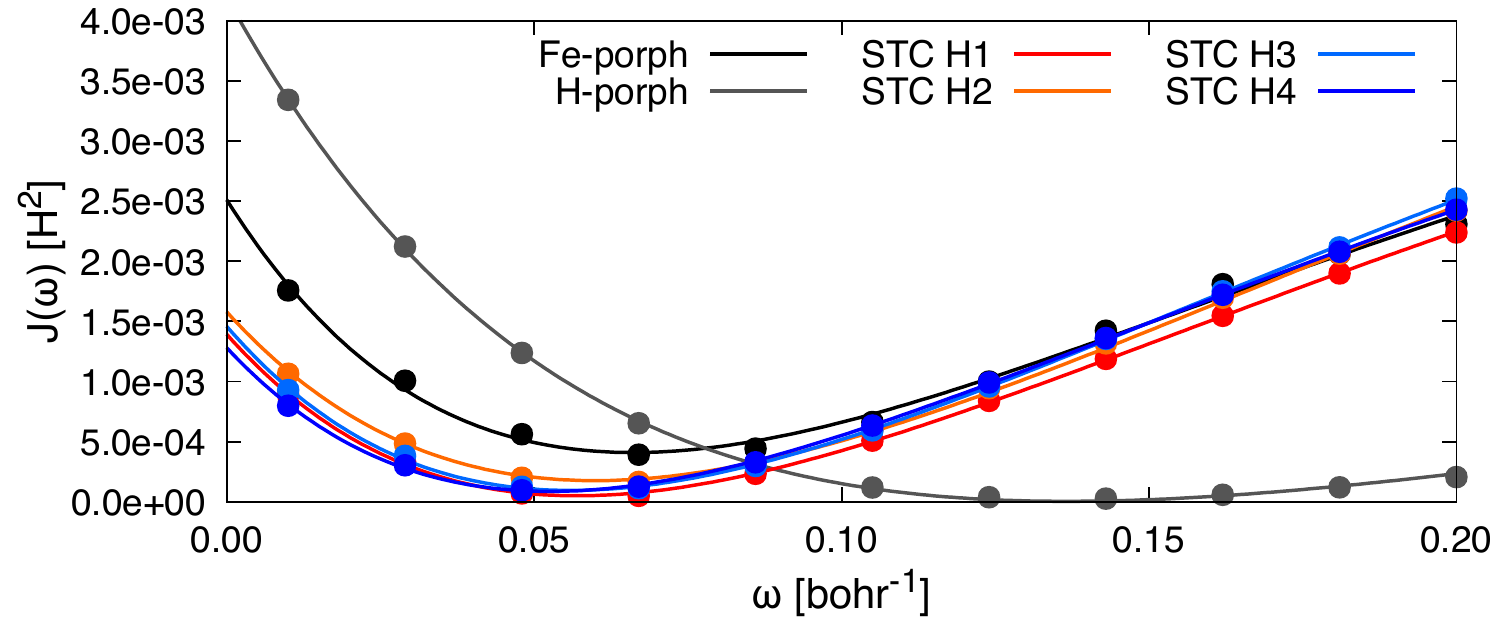}
  \caption{
Dependence of penalty function $J$, Eq.~\ref{eq_otrsh_j} on range-separation parameter $\omega$ for symmetric bis-histidine coordinated Fe$^{3+}$-porphine (Fe-porph), for the same porphine with Fe substituted by two H atoms (H-porph) and for the bis-histidine coordinated Fe$^{3+}$-porphyrines with geometries taken from the Au-STC-Au junction model structure (STC H1 to STC H4). 
The calculated points are fitted by Morse potential-type curves.}
  \label{fig_otrsh_omega}
\end{figure}

Here we optimized $\omega$  for the 4 bis-histidine coordinated Fe$^{3+}$-porphines taking the geometry from the Au-STC-Au junction model structure (all heme side chains were replaced by hydrogens and the axial histidines were modelled by imidazoles) as well as for idealised symmetric bis-histidine Fe$^{3+}$-porphine  
(used in work of Smith \emph{et al.}~\cite{Smith2003, Smith2006}).
For the latter, structure with the central Fe cation replaced by 2 hydrogen atoms was evaluated as well.
The PBE\cite{Perdew1996} functional is used for the GGA part. 
The total electronic energy of Fe$^{3+}$-heme, Fe$^{4+}$-heme and Fe$^{2+}$-porphyrines was calculated for the experimental spin ground states, doublet, triplet and singlet, respectively.
The dependence of $J$ on $\omega$ is shown in Fig.~\ref{fig_otrsh_omega} and numerical values for $\omega$ are listed in Table~\ref{tab_band_gap} together with band gaps obtained at OT-RSH and PBE levels and the corresponding correction $\Sigma_0^\alpha$ for HOMO and LUMO one-electron energies.

Interestingly, the range separation distances are rather long for the heme cofactors (9-10 \AA) compared to 3-4 \AA\ typically found for organic molecules like transition metal-free porphins. 
Because of the long separation distances, good performance of global hybrid functionals can be expected for these molecules. 
We find that the HOMO level decreases by $\sim$1.3 eV and the LUMO level shifts up by $\sim$1.5 eV compared to PBE resulting in an increase of the HOMO-LUMO energy gap from 0.7 in PBE to 3.5 eV at OT-RSH level. 


\subsection{Interfacial state renormalization}

\begin{figure}[thb]
  \centering
  \includegraphics[scale=1.0]{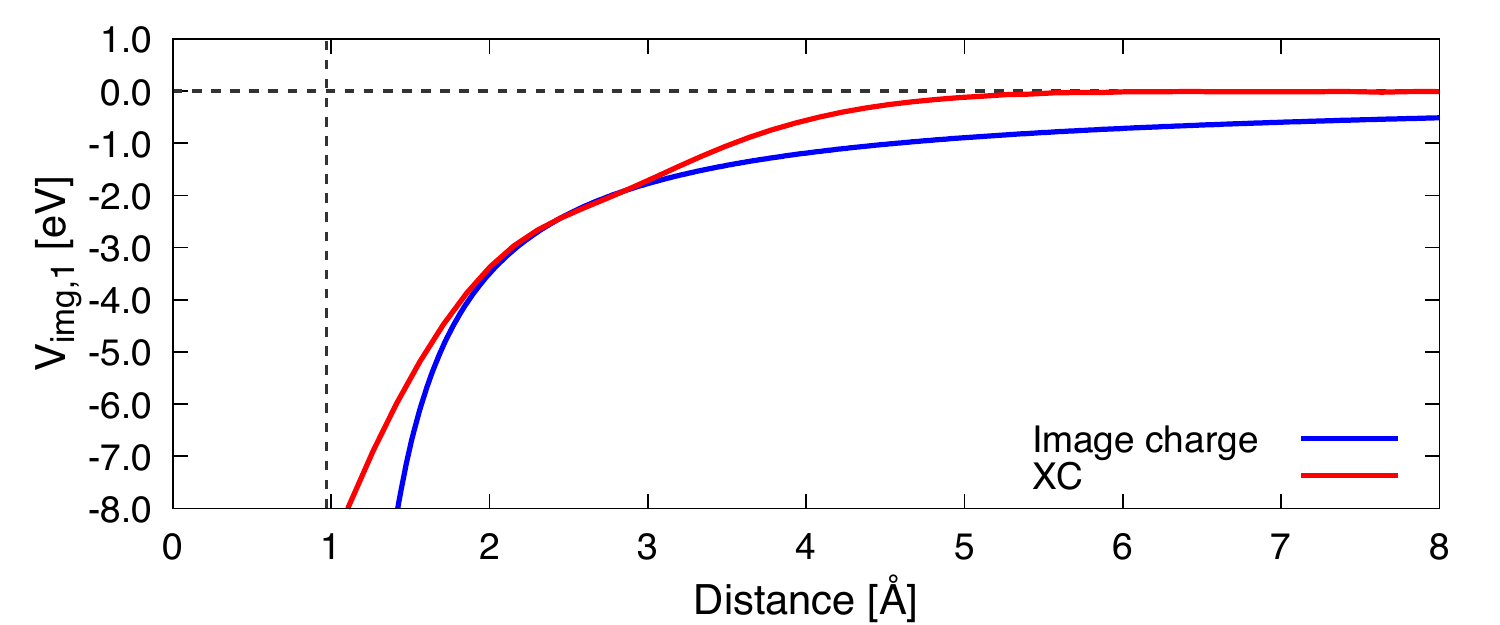}
  \caption{
Classical image charge potential Eq.~\ref{eq_img_pot_1} ($q\!=\!1$) fitted to the exchange-correlation (XC) potential calculated using DFT at PBE level on 2-layer Au(111) vacuum surface.
The surface-plane position obtained is located at $z_0 = 0.97$ \AA.}
  \label{fig_xc_pot}
\end{figure}

The interaction between the metal and the occupied and unoccupied states of the molecule gives rise to electronic polarization of the metal and a change of the 
molecular orbital energies. 
This leads to a reduction of the band gap, which is called state renormalization. 
Following the work of Neaton \emph{et al.}~\cite{Neaton2006} we calculate the polarization energy from a classical image-charge potential, 
\begin{equation}
  V_\mathrm{img,1}(z) = 
    -\frac{q}{4 \pi \epsilon_0} 
     \frac{1}{4|z - z_0|} 
  \label{eq_img_pot_1} 
\end{equation}
that is fitted to the DFT exchange-correlation (XC) potential above the gold surface at vacuum (no molecule present), see Fig.~\ref{fig_xc_pot}. 
The XC potential was calculated for the two-monolayer Au(111) slab in our DFT model using the same cell size and computational parameters as for the electronic structure calculations of the protein junction (see Section~\ref{sec_dft_model}). 
The parameter $z_0$ of the image-charge potential Eq.~\ref{eq_img_pot_1} was obtained by finding the common tangent point with the XC potential, giving $z_0$ = 0.97 \AA.

For the case of two gold surfaces, as relevant for the Au-STC-Au junction, the image charge potential was described by a converging series 
\begin{equation}
  V_\mathrm{img,2}(z) = 
    \frac{q}{4\pi\epsilon_0} \sum_{n=-\infty}^\infty
    \left( \frac{1}{|z-2nd-z_0|} - \frac{1}{|z-2nd+z_0|} \right),  
  \label{eq_img_pot_2} 
\end{equation}
where $q$ is the charge of the particle and $d$ is the distance between the two surface planes. 
Eq.~\ref{eq_img_pot_2} was applied to calculate the junction potential using the gold-separation distance $d$ = 26.8~\AA\ from the DFT model and $z_0$ = 0.97~\AA\ from the $V_{img,1}$ fit to the surface XC potential.

Since the hemes/protein are not parallel to the surfaces, the question arises how one should distribute the charge $q$ over the protein for calculation of the image charge potential. 
In a first approximation we calculate the image charge potential for each heme-cofactor separately and then average $\Sigma^{\text{HOMO}}$ and $\Sigma^{\text{LUMO}}$ over all four cofactors to obtain an effective shift for occupied and unoccupied levels. 
At first we placed the unit charge at the positions of the Fe ions of the heme cofactors. 
The distance from the left and right surfaces in the direction of the surface normal ($d_L$ and $d_R$), the calculated image-charge energies ($\Sigma_{pol}^\alpha$) and the total corrections $\Sigma^\alpha\!=\! \Sigma_0^\alpha + \Sigma_{pol}^{\alpha}$ are summarized in Table~\ref{tab_band_renorm_fe}. 
The average polarization energy was found to be $\Sigma_{pol}^{\text{HOMO}}\!=\!-\Sigma_{pol}^{\text{LUMO}} \sim 0.9$ eV predicting a total shift $\Sigma^{\text{HOMO}} \sim -0.4$ eV and $\Sigma^{\text{LUMO}}\sim 0.6$ eV.  
Next we considered a more realistic charge distribution on the heme cofactors and calculated the atomic point-charges by Mulliken~\cite{Mulliken1955} and Hirshfeld~\cite{Hirshfeld1977} population analyses.
Atomic point charges were calculated for the HOMO and LUMO orbitals of the bis-histidine coordinated Fe$^{3+}$-porphine models with  geometries taken from the structural model of the Au-STC-Au junction, as before.
The calculated image-charge energies and the corresponding total corrections for HOMO and LUMO are collected in Table~\ref{tab_band_renorm_chrg}.
Now the HOMO level is shifted more significantly, $\Sigma^{\text{HOMO}} \sim -0.8$ eV on average, with respect to their PBE positions. 
The LUMO correction remains similar as before, $\Sigma^{\text{HOMO}} \sim 0.7$ eV. 
The polarization energies differ between Mulliken and Hirshfeld charge on individual heme cofactors, however, they are similar on average.

Finally, in our most realistic model, we consider the charge distribution over the actual protein HOMO and LUMO orbitals as obtained from DFT/POD calculations on the full Au-STC-Au junction. 
Hence, instead of calculating a correction for each individual heme-cofactor model and averaging, now a single correction is calculated for the actual HOMO and LUMO orbitals of the protein. 
The latter are delocalized mostly over the four heme-cofactors (94\%) and only a small fraction is delocalized over the protein (6\%).
Hence, the image-charge correction was calculated as 
\begin{equation}
  \Sigma_{pol}^{\alpha} = q \int_V 
    |\psi^\alpha(\mathbf{r})|^2 V_\mathrm{img}(\mathbf{r}) d\mathrm{r}^3
\end{equation}
where $\psi^\alpha$ is the normalized frontier orbital, $\alpha\!=\!$ HOMO or LUMO and charge $q\!=\!1$ for HOMO and $q\!=\!-1$ for the LUMO, respectively.
The integration was done numerically on the grid using a grid spacing of $\sim$0.15 \AA\ in each direction.
The image-charge corrections as well as the total HOMO and LUMO corrections are listed in Table~\ref{tab_band_renorm_mo}.
The total correction for the HOMO is $-1.2$ eV while that for the LUMO is 1.4 eV with respect to the PBE values. 
Therefore, according to our best DFT+$\Sigma$ estimates, the top of the valence band and the bottom of the conduction band are at -1.2~eV and 1.6 eV with respect to the Fermi level of the electrode (at zero voltage). 
The former value compares favourably with the experimental estimate, $-1.2$ eV, as determined from photo-emission spectroscopy and work function measurement. 
We also conclude that the single point-charge approach underestimates the correction while the all-atomic point-charge approach fails to predict qualitatively correct shift of the HOMO level because of over-polarization of the hydrogen atoms near the gold surface.
On the other hand, the MO-integration approach is perfect agreement with the UPS experimental measurements of the HOMO level position.

\clearpage



\begin{table}[tbh]
  \centering
  \caption{Parameters of the incoherent hopping model: mean heme-pair electronic coupling $\langle|H_{ij}|^2\rangle^{1/2}$ [meV], heme-pair Fe-Fe distance along the direction from the first to the last site $r_{ij}$ [\AA].}
  \begin{tabular}{ccc}
    \hline
    Pair & 
    $\langle|H_{ij}|^2\rangle^{1/2}$ &
    $r_{ij}$ \\
    \hline
    1-2  & 2.17 & 3.89 \\
    2-3  & 3.08 & 5.37 \\
    3-4  & 2.08 & 5.39 \\ 
    \hline
  \end{tabular}
  \label{tab_hopping}
\end{table}

\begin{table}[tbh]
  \centering
  \caption{Correction for the energy gap in vacuum for bis-histidine coordinated Fe$^{3+}$-porphyrines. 
 The gap was calculated on geometries extracted from the model structure of the Au-STC-Au junction used for the I-V calculations. 
 The range-separation parameter $\omega$ is given in bohr$^{-1}$, corresponding to a range-separation distance 
 $\rho = 1/\omega$ in \AA. $E_g$ is the HOMO-LUMO energy gap and $\Sigma_0^\mathrm{HOMO}$ and $\Sigma_0^\mathrm{LUMO}$ 
 are the energy level corrections for the HOMO and LUMO in eV. 
  }
  \vspace{1ex}
  \begin{tabular}{lccccccc}
    \hline
    Cofactor & ResID & 
    $\omega$ & $\rho$ & 
    $E_g^\mathrm{PBE}$ &
    $E_g^\mathrm{OT-RSH}$ &
    $\Sigma_0^\mathrm{HOMO}$ &
    $\Sigma_0^\mathrm{LUMO}$ \\
    \hline
    Ideal (with Fe) & --- 
      &  0.066 &  8.01 & 0.548 & 3.922 & -1.527 & 1.847 \\
    Ideal (w/o Fe)  & --- 
      &  0.137 &  3.87 & 0.485 & 4.155 & -2.035 & 1.635 \\
    \hline
    Heme 1 & 92
      &  0.057 &  9.29 & 0.641 & 3.463 & -1.298 & 1.523 \\
    Heme 2 & 94
      &  0.056 &  9.48 & 0.751 & 3.680 & -1.360 & 1.568 \\
    Heme 3 & 95
      &  0.053 & 10.08 & 0.700 & 3.292 & -1.190 & 1.402 \\
    Heme 4 & 93
      &  0.060 &  8.83 & 0.735 & 3.598 & -1.258 & 1.605 \\
    \hline
    Average & --- 
      &  0.056 &  9.42 & 0.707 & 3.508 & -1.276 & 1.525 \\
    \hline
  \end{tabular}
  \label{tab_band_gap}
\end{table}

\begin{table}[tbh]
  \centering
  \caption{Interfacial band renormalization determinated by image-charge interaction of the positive unit charge located at the iron position of the each heme cofactor. Distance of the charge from the left ($d_L$) and right ($d_R$) electrode are given in \AA\ while the image-charge energy $\Sigma_{pol}$ and final HOMO/LUMO corrections $\Sigma^\mathrm{HOMO}$, $\Sigma^\mathrm{LUMO}$ are given in eV.}
  \vspace{1ex}
  \begin{tabular}{lccccccc}
    \hline
    Cofactor & ResID & $d_L$ & $d_R$ & 
    $\Sigma_{pol}^\mathrm{HOMO}$ &
    $\Sigma_{pol}^\mathrm{LUMO}$ &
    $\Sigma^\mathrm{HOMO}$ &
    $\Sigma^\mathrm{LUMO}$ \\
    \hline
    Heme 1 & 92 
      & 16.94 &  9.84 & 0.648 & -0.648 & -0.650 & 0.875 \\
    Heme 2 & 94
      &  7.37 & 19.41 & 1.113 & -1.113 & -0.247 & 0.455 \\
    Heme 3 & 95 
      &  9.50 & 17.28 & 0.694 & -0.694 & -0.496 & 0.708 \\
    Heme 4 & 94
      & 19.48 &  7.30 & 1.133 & -1.133 & -0.125 & 0.472 \\
    \hline
    Average & ---
      &  ---  &  ---  & 0.897 & -0.897 & -0.379 & 0.628 \\
    \hline
  \end{tabular}
  \label{tab_band_renorm_fe}
\end{table}

\begin{table}[tbh]
  \centering
  \caption{Interfacial band renormalization determined by image-charge interaction of atomic point-charges on heme-cofactor atoms in oxidized state. Distance of the heme from the left ($d_L$) and right ($d_R$) electrode are given in \AA\ while the image-charge energy $\Sigma_{pol}$ and final HOMO/LUMO corrections $\Sigma^\mathrm{HOMO}$, $\Sigma^\mathrm{LUMO}$ are given in eV.}
  \vspace{1ex}
  \setlength{\tabcolsep}{4pt}
  \begin{tabular}{lcccccccccc}
    \hline
    \multirow{2}{*}{Heme} & 
    \multirow{2}{*}{$d_L$} & 
    \multirow{2}{*}{$d_R$} & 
    \multicolumn{4}{c}{Mulliken charges} &
    \multicolumn{4}{c}{Hirshfeld charges} \\
    & & &
    $\Sigma_{pol}^\mathrm{HOMO}$ &
    $\Sigma_{pol}^\mathrm{LUMO}$ &
    $\Sigma^\mathrm{HOMO}$ &
    $\Sigma^\mathrm{LUMO}$ &
    $\Sigma_{pol}^\mathrm{HOMO}$ &
    $\Sigma_{pol}^\mathrm{LUMO}$ &
    $\Sigma^\mathrm{HOMO}$ &
    $\Sigma^\mathrm{LUMO}$ \\
    \hline
     1 & 11.69 &  4.72 & 0.854 & -0.691 & -0.444 & 0.832 & 
                         0.948 & -0.661 & -0.350 & 0.862 \\
     2 &  3.21 & 15.01 & 2.286 & -0.809 &  0.926 & 0.759 & 
                         2.747 & -0.520 &  1.387 & 1.048 \\
     3 &  5.28 & 12.72 & 0.807 & -0.713 & -0.383 & 0.689 & 
                         0.861 & -0.703 & -0.329 & 0.699 \\
     4 & 14.60 &  3.62 & 1.900 & -0.975 &  0.642 & 0.630 & 
                         2.225 & -0.766 &  0.967 & 0.839 \\
    \hline
     Avrg
       &  ---  &  ---  & 1.462 & -0.797 &  0.185 & 0.728 &
                         1.695 & -0.663 &  0.419 & 0.862 \\
    \hline
  \end{tabular}
  \label{tab_band_renorm_chrg}
\end{table}

\begin{table}[tbh]
  \centering
  \caption{Interfacial band renormalization determinated by image-charge interaction of frontier molecular orbitals of the oxidized Au/STC junctions. Distance of the heme from the left ($d_L$) and right ($d_R$) electrode are given in \AA\ while the image-charge energy $\Sigma_{pol}$ and final HOMO/LUMO corrections $\Sigma^\mathrm{HOMO}$, $\Sigma^\mathrm{LUMO}$ are given in eV.}
  \vspace{1ex}
  \begin{tabular}{cccc}
    \hline
    $\Sigma_{pol}^\mathrm{HOMO}$ &
    $\Sigma_{pol}^\mathrm{LUMO}$ &
    $\Sigma^\mathrm{HOMO}$ &
    $\Sigma^\mathrm{LUMO}$ \\
    \hline
      0.031 & -0.113 & -1.245 & 1.412 \\
    \hline
  \end{tabular}
  \label{tab_band_renorm_mo}
\end{table}
\clearpage

\begin{table}[tbh]
  \centering
  \caption{
The first 10 most contributing states $j$ (conduction channels) to the toal current $I_0$ at bias voltage 0.5~V in normal, off-resonant tunneling regime as well as in hypothetical resonant regime.
State energies with respect to Fermi level $E_F$ are given in eV. 
$P_\mathrm{heme}$, $P_\mathrm{Fe}$ and $P_\mathrm{aa}^\mathrm{surf}$ denotes projection (localization) of the given state on the bis-histidine hemes (without Fe), the iron atoms and the surface amino acids listed in Table~\ref{tab_surf_aa}, respectively.}
  \vspace{1ex}
  \setlength{\tabcolsep}{4pt}
  \begin{tabular}{c@{\hspace{10mm}}cccccc@{\hspace{10mm}}cccccc}
    \hline
      \multirow{2}{*}{N} & 
      \multicolumn{6}{c}{off-resonant regime} &
      \multicolumn{6}{c}{resonant regime} \\
      & State & $I_j/I_0$ & $E-E_F$ & 
        $P_\mathrm{heme}$ & $P_\mathrm{Fe}$ & $P_\mathrm{aa}^\mathrm{surf}$
      & State & $I_j/I_0$ & $E-E_F$ & 
        $P_\mathrm{heme}$ & $P_\mathrm{Fe}$ & $P_\mathrm{aa}^\mathrm{surf}$ \\
    \hline
       1 & 2067 & 0.088 & -3.007 & 0.136 & 0.017 & 0.371 &
           2264 & 0.522 & -0.295 & 0.528 & 0.223 & 0.238 \\
       2 & 2066 & 0.044 & -3.010 & 0.114 & 0.004 & 0.382 & 
           2266 & 0.313 & -0.289 & 0.397 & 0.372 & 0.213 \\
       3 & 2167 & 0.027 & -2.375 & 0.526 & 0.005 & 0.328 &
           2067 & 0.019 & -1.807 & 0.137 & 0.017 & 0.371 \\
       4 & 2058 & 0.026 & -3.051 & 0.088 & 0.013 & 0.301 &
           2066 & 0.009 & -1.810 & 0.114 & 0.004 & 0.382 \\
       5 & 2057 & 0.024 & -3.057 & 0.066 & 0.010 & 0.708 &
           2167 & 0.009 & -1.175 & 0.526 & 0.005 & 0.328 \\
       6 & 2168 & 0.016 & -2.373 & 0.192 & 0.001 & 0.604 &
           2267 & 0.007 & -0.286 & 0.450 & 0.540 & 0.004 \\
       7 & 2068 & 0.016 & -3.004 & 0.198 & 0.008 & 0.463 &
           2168 & 0.005 & -1.173 & 0.192 & 0.001 & 0.604 \\
       8 & 1941 & 0.015 & -4.011 & 0.357 & 0.008 & 0.500 &
           2058 & 0.005 & -1.851 & 0.088 & 0.013 & 0.301 \\
       9 & 2264 & 0.014 & -1.495 & 0.528 & 0.223 & 0.238 &
           2057 & 0.005 & -1.857 & 0.066 & 0.010 & 0.708 \\
      10 & 1765 & 0.014 & -5.097 & 0.081 & 0.002 & 0.725 &
           2263 & 0.005 & -0.309 & 0.054 & 0.006 & 0.702 \\
    \hline
  \end{tabular}
  \label{tab_iv_cont}
\end{table}

\begin{table}[tbh]
  \centering
  \caption{
Projection of the conduction channels (MOs) from Table~\ref{tab_iv_cont} to the surface amino acids within the van der Waals distance (up to 3.5\AA) from a gold surface. 
For comparison, localization on the heme cofactors is shown as well.
Cysteines marked by star symbols ($*$) are covalently bound to the hemes while the hash-marked ($\#$) cysteine 87 is covalently bound to the gold surface.
Axial histidines coordinated to Fe cations are involved in the hemes.
ACE stands for N-terminal acetyl while NHE means C-terminal NH$_2$ group.
Shortest distances to the left ($R_L$) and right ($R_R$) electrodes are given \AA\ with the contact atom name specified in Amber force-field format.}
  \vspace{1ex}
  \setlength{\tabcolsep}{4pt}
  \begin{tabular}{cccccccccccccc}
    \hline
      \multirow{2}{*}{Residue} & 
      \multirow{2}{*}{Atom} & 
      \multirow{2}{*}{$R_L$} & 
      \multirow{2}{*}{$R_R$} &
      \multicolumn{10}{c}{Molecular orbital} \\
       & & & &
      2067 & 2066 & 2167 & 2058 & 2057 &
      2168 & 2068 & 1941 & 2264 & 1765 \\
    \hline
      Heme 1      & HP74 & 11.69 &  4.72 &
        0.004 & 0.001 & 0.000 & 0.000 & 0.000 &
        0.000 & 0.000 & 0.023 & 0.000 & 0.003 \\
      Heme 2      & HM52 &  3.21 & 15.01 &
        0.046 & 0.095 & 0.012 & 0.006 & 0.000 &
        0.002 & 0.170 & 0.001 & 0.002 & 0.029 \\
      Heme 3      & O1A  &  5.28 & 12.72 &
        0.017 & 0.002 & 0.468 & 0.079 & 0.068 &
        0.150 & 0.008 & 0.011 & 0.177 & 0.021 \\
      Heme 4      & HM53 & 14.60 &  3.62 &
        0.086 & 0.021 & 0.050 & 0.017 & 0.007 &
        0.041 & 0.005 & 0.353 & 0.571 & 0.029 \\
    \hline
      ACE-1       & O    &  2.65 & 21.85 & 
        0.000 & 0.000 & 0.000 & 0.000 & 0.000 &
        0.000 & 0.000 & 0.000 & 0.000 & 0.000 \\
      Glu-3       & HA   &  2.99 & 20.95 &
        0.000 & 0.000 & 0.000 & 0.000 & 0.000 &
        0.000 & 0.000 & 0.000 & 0.000 & 0.001 \\
      Phe-8       & HB2  &  3.03 & 20.99 &
        0.000 & 0.000 & 0.000 & 0.000 & 0.000 &
        0.000 & 0.000 & 0.000 & 0.000 & 0.001 \\
      Glu-11      & HN   &  2.55 & 21.13 &
        0.000 & 0.000 & 0.000 & 0.000 & 0.000 &
        0.000 & 0.000 & 0.000 & 0.000 & 0.000 \\
      Ser-12      & HN   &  2.59 & 21.82 &
        0.000 & 0.000 & 0.000 & 0.000 & 0.000 &
        0.000 & 0.000 & 0.000 & 0.000 & 0.000 \\
      Ser-37      & O    &  3.20 & 19.28 &
        0.009 & 0.013 & 0.000 & 0.002 & 0.000 &
        0.000 & 0.051 & 0.000 & 0.000 & 0.000 \\
      Cys$^*$-38  & O    &  2.37 & 21.06 &
        0.046 & 0.067 & 0.000 & 0.005 & 0.000 &
        0.000 & 0.266 & 0.000 & 0.000 & 0.001 \\
      Gly-40      & HN   &  2.81 & 20.85 &
        0.001 & 0.000 & 0.000 & 0.001 & 0.002 &
        0.000 & 0.002 & 0.000 & 0.003 & 0.002 \\
      Lys-41      & NZ   &  2.67 & 19.87 &
        0.001 & 0.000 & 0.000 & 0.003 & 0.009 &
        0.000 & 0.000 & 0.000 & 0.000 & 0.000 \\
      Glu-44      & HG2  &  2.64 & 20.55 &
        0.001 & 0.000 & 0.001 & 0.025 & 0.236 &
        0.001 & 0.002 & 0.003 & 0.000 & 0.005 \\
      Met-45      & HE2  &  2.58 & 20.90 &
        0.000 & 0.000 & 0.010 & 0.039 & 0.366 &
        0.014 & 0.000 & 0.016 & 0.122 & 0.006 \\
      Asp-46      & HA   &  2.89 & 19.40 &
        0.000 & 0.000 & 0.069 & 0.005 & 0.035 &
        0.080 & 0.000 & 0.002 & 0.001 & 0.002 \\
      Ala-47      & HB2  &  3.04 & 19.98 &
        0.000 & 0.000 & 0.014 & 0.004 & 0.017 &
        0.016 & 0.000 & 0.000 & 0.000 & 0.004 \\
      Lys-50      & HZ2  &  2.34 & 16.32 &
        0.000 & 0.000 & 0.020 & 0.000 & 0.001 &
        0.015 & 0.000 & 0.001 & 0.000 & 0.447 \\
      Val-69      & HG21 &  2.69 & 20.66 &
        0.005 & 0.057 & 0.000 & 0.001 & 0.000 &
        0.000 & 0.028 & 0.000 & 0.000 & 0.007 \\
      Gly-70      & HA2  &  2.96 & 22.20 &
        0.001 & 0.006 & 0.000 & 0.000 & 0.000 &
        0.000 & 0.003 & 0.000 & 0.000 & 0.001 \\
      Lys-72      & HZ1  &  2.28 & 16.76 &
        0.026 & 0.161 & 0.002 & 0.009 & 0.001 &
        0.001 & 0.001 & 0.000 & 0.000 & 0.057 \\
      Cys$^\#$-87 & SG   &  2.22 & 20.45 &
        0.000 & 0.000 & 0.000 & 0.000 & 0.000 &
        0.000 & 0.000 & 0.000 & 0.000 & 0.003 \\
      Val-88      & HA   &  3.34 & 20.53 &
        0.000 & 0.000 & 0.000 & 0.000 & 0.000 &
        0.000 & 0.000 & 0.000 & 0.000 & 0.001 \\
      Leu-89      & HD13 &  2.70 & 20.10 &
        0.000 & 0.000 & 0.000 & 0.000 & 0.000 &
        0.000 & 0.000 & 0.000 & 0.000 & 0.003 \\
      Lys-90      & HZ2  &  1.96 & 20.98 &
        0.000 & 0.000 & 0.000 & 0.000 & 0.000 &
        0.000 & 0.000 & 0.000 & 0.000 & 0.052 \\
      Lys-91      & HZ2  &  2.52 & 20.51 &
        0.000 & 0.000 & 0.001 & 0.000 & 0.000 &
        0.001 & 0.000 & 0.000 & 0.000 & 0.003 \\
      NHE-92      & HN1  &  1.98 & 24.02 &
        0.000 & 0.000 & 0.000 & 0.000 & 0.000 &
        0.000 & 0.000 & 0.000 & 0.000 & 0.011 \\
    \hline
      Pro-24      & HB1  & 21.14 &  2.66 &
        0.000 & 0.001 & 0.000 & 0.000 & 0.000 &
        0.000 & 0.001 & 0.000 & 0.000 & 0.000 \\
      Ser-25      & HG1  & 21.19 &  2.73 &
        0.000 & 0.000 & 0.000 & 0.000 & 0.000 &
        0.000 & 0.000 & 0.000 & 0.000 & 0.000 \\
      Ala-26      & HB3  & 21.45 &  2.60 &
        0.000 & 0.000 & 0.000 & 0.000 & 0.000 &
        0.000 & 0.000 & 0.000 & 0.000 & 0.001 \\
      Asp-27      & HB2  & 21.04 &  2.62 &
        0.000 & 0.001 & 0.000 & 0.001 & 0.000 &
        0.000 & 0.002 & 0.000 & 0.000 & 0.000 \\
      Phe-30      & HZ   & 15.85 &  2.69 &
        0.002 & 0.005 & 0.000 & 0.002 & 0.000 &
        0.000 & 0.013 & 0.000 & 0.000 & 0.004 \\
      Ser-77      & O    & 21.61 &  2.45 &
        0.054 & 0.021 & 0.039 & 0.035 & 0.006 &
        0.078 & 0.001 & 0.174 & 0.001 & 0.001 \\
      Cys$^*$-78  & O    & 20.88 &  2.82 &
        0.049 & 0.017 & 0.154 & 0.010 & 0.002 &
        0.317 & 0.001 & 0.274 & 0.107 & 0.008 \\
      Asp-80      & HB1  & 21.27 &  2.61 &
        0.069 & 0.010 & 0.002 & 0.154 & 0.031 &
        0.002 & 0.003 & 0.010 & 0.000 & 0.006 \\
      Asp-81      & HB2  & 21.08 &  2.71 &
        0.005 & 0.001 & 0.010 & 0.001 & 0.000 &
        0.004 & 0.000 & 0.019 & 0.000 & 0.005 \\
      Arg-83      & HH22 & 16.59 &  3.33 &
        0.101 & 0.021 & 0.004 & 0.003 & 0.000 &
        0.074 & 0.007 & 0.001 & 0.003 & 0.013 \\
    \hline
  \end{tabular}
  \label{tab_surf_aa}
\end{table}

\clearpage

\begin{figure}[thb]
  \centering
  \begin{tabular}{cc}
    \includegraphics[width=65mm,height=35mm]{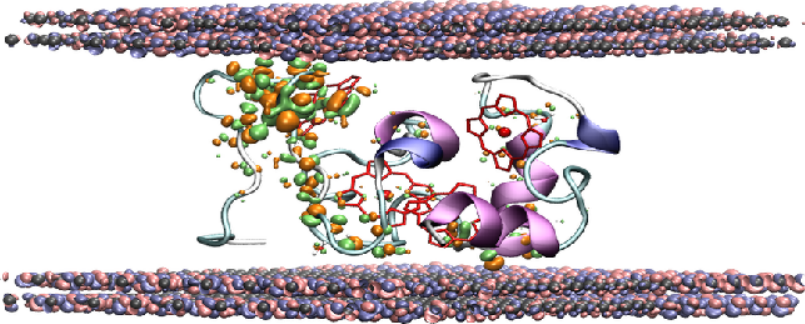}(a) &
    \includegraphics[width=65mm,height=35mm]{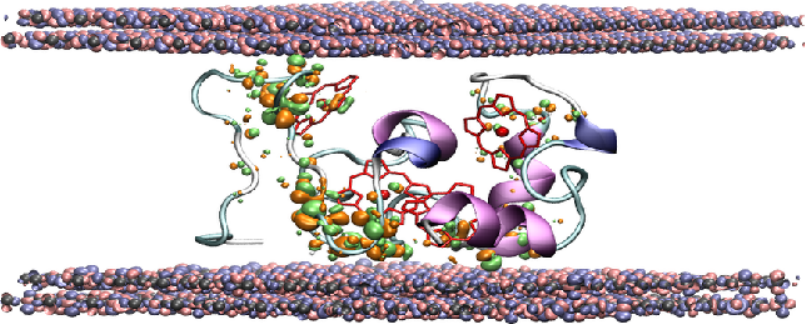}(b) \\
    \includegraphics[width=65mm,height=35mm]{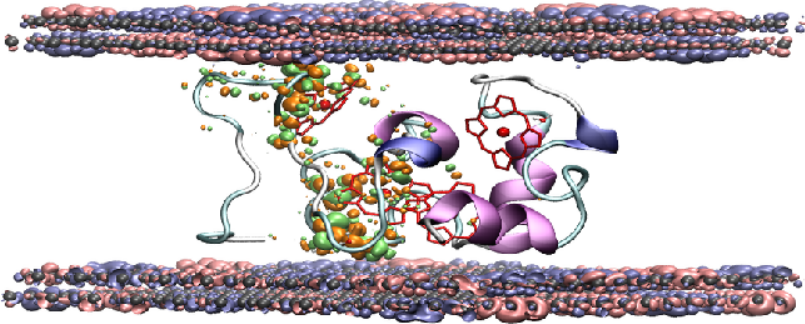}(c) &
    \includegraphics[width=65mm,height=35mm]{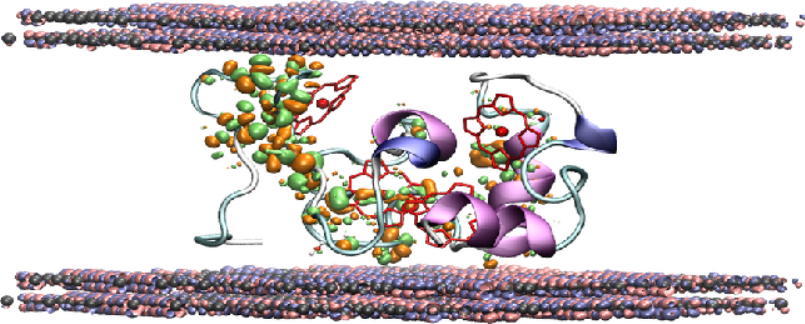}(d) \\
    \includegraphics[width=65mm,height=35mm]{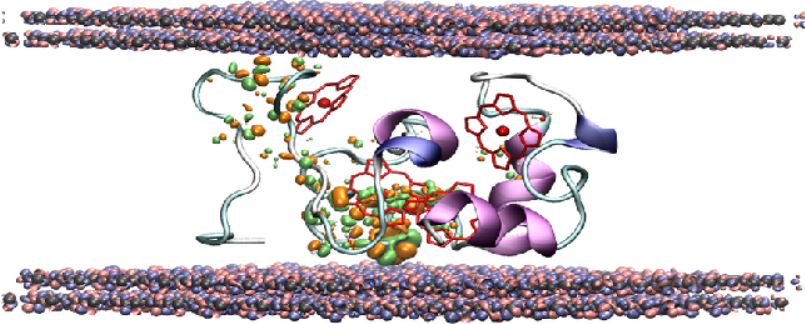}(e) &
    \includegraphics[width=65mm,height=35mm]{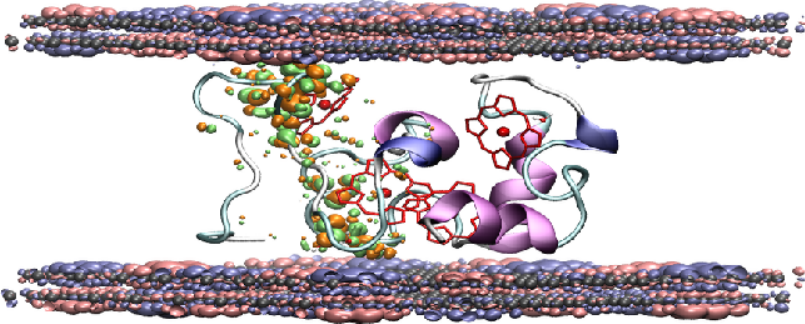}(f) \\
    \includegraphics[width=65mm,height=35mm]{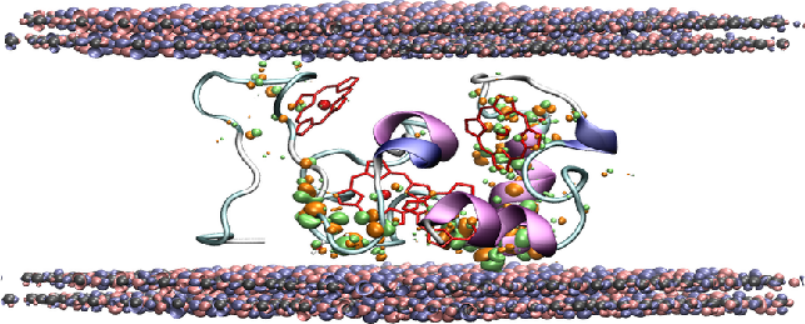}(g) &
    \includegraphics[width=65mm,height=35mm]{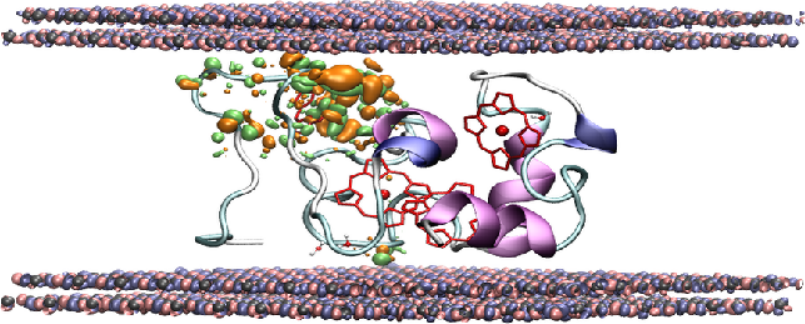}(h) \\
    \includegraphics[width=65mm,height=35mm]{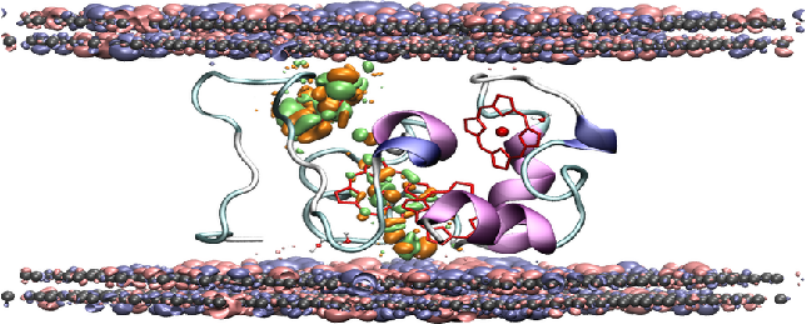}(i) &
    \includegraphics[width=65mm,height=35mm]{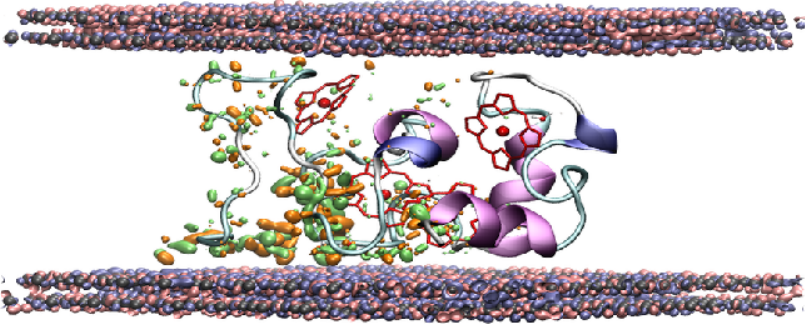}(j) \\
  \end{tabular}
  \caption{
First 10 dominant conduction channels in the Au-STC-Au junction. The top of the valence band was aligned with the experimentally determined position as explained in the main text. Position and current contributions of the channels are 
(a) -3.007 eV, 8.8\%; 
(b) -3.010 eV, 4.4\%; 
(c) -2.375 eV, 2.7\%; 
(d) -3.051 eV, 2.6\%; 
(e) -3.057 eV, 2.4\%; 
(f) -2.373 eV, 1.6\%; 
(g) -3.004 eV, 1.6\%; 
(h) -4.011 eV, 1.5\%; 
(i) -1.495 eV, 1.4\%; 
(j) -5.097 eV, 1.4\%. 
The channels are described in Table~\ref{tab_iv_cont}.}
  \label{fig_channel_off}
\end{figure}

\clearpage

\begin{figure}[thb]
  \centering
  \begin{tabular}{cc}
    \includegraphics[width=65mm,height=35mm]{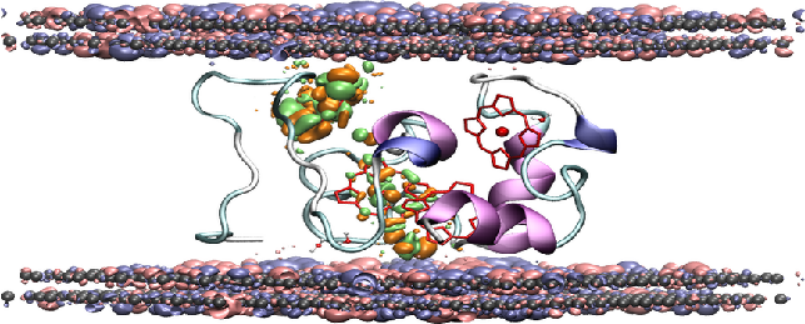}(a) &
    \includegraphics[width=65mm,height=35mm]{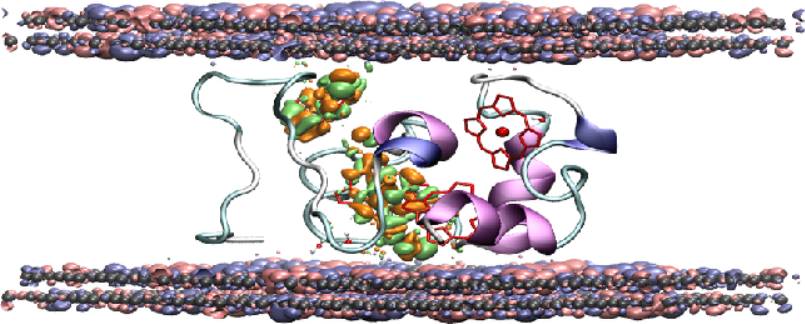}(b) \\
    \includegraphics[width=65mm,height=35mm]{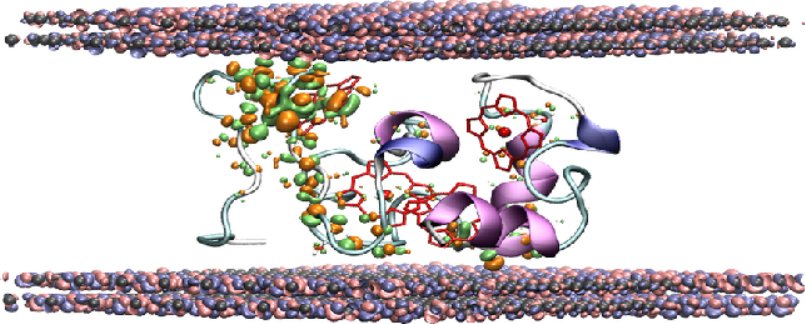}(c) &
    \includegraphics[width=65mm,height=35mm]{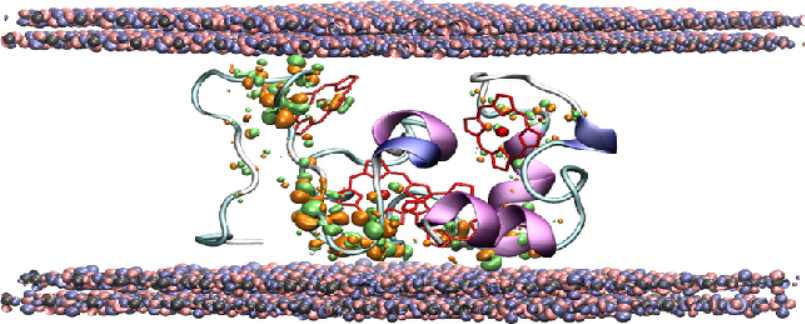}(d) \\
    \includegraphics[width=65mm,height=35mm]{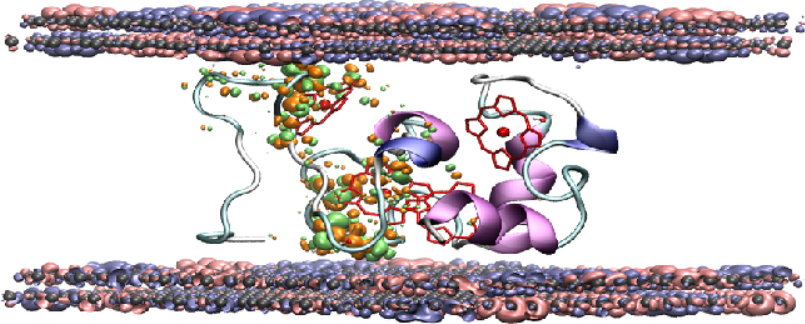}(e) &
    \includegraphics[width=65mm,height=35mm]{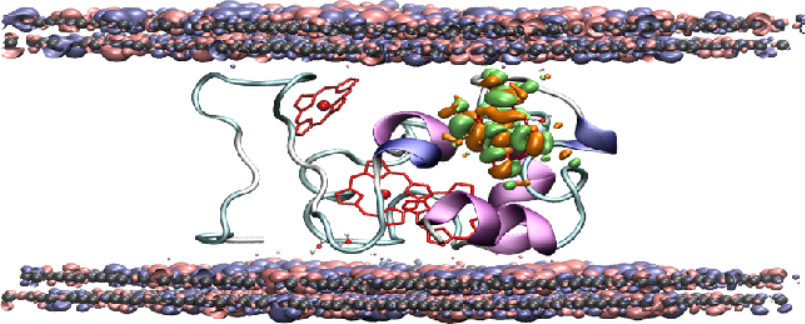}(f) \\
    \includegraphics[width=65mm,height=35mm]{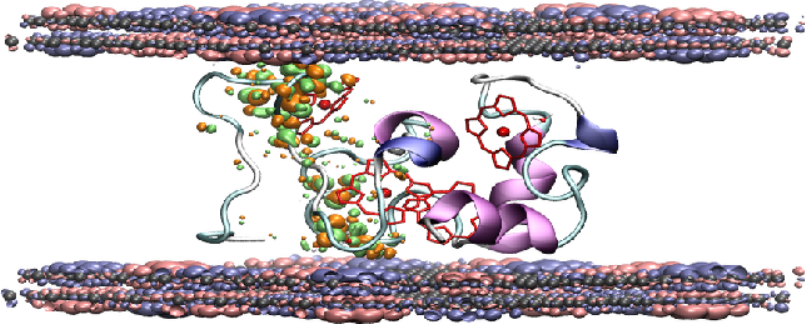}(g) &
    \includegraphics[width=65mm,height=35mm]{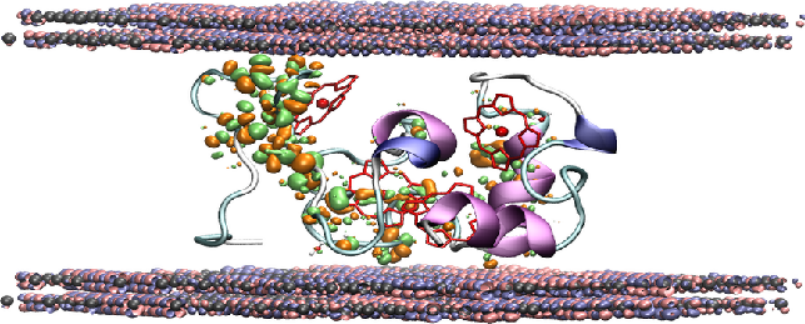}(h) \\
    \includegraphics[width=65mm,height=35mm]{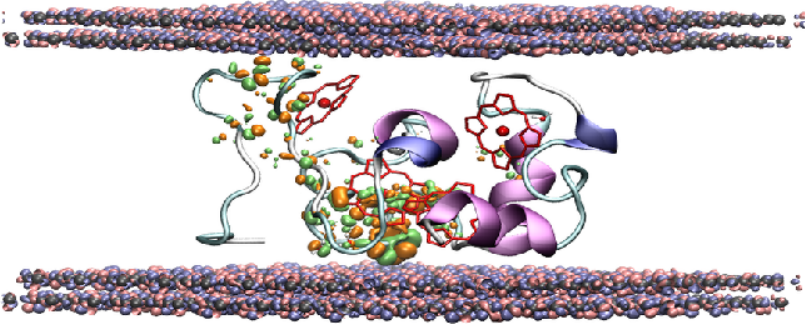}(i) &
    \includegraphics[width=65mm,height=35mm]{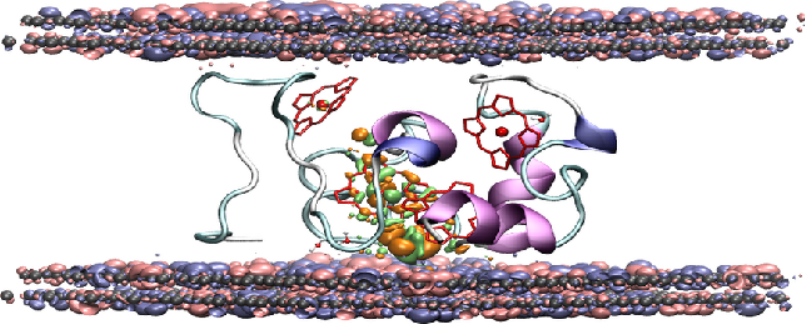}(j) \\
  \end{tabular}
  \caption{
First 10 dominant conduction channels in the Au-STC-Au junction. The top of the valence band was aligned with the Fermi level to simulate hypothetical resonant regime. Position and current contributions of the channels are 
(a) -0.295 eV, 52.2\%; 
(b) -0.289 eV, 31.3\%; 
(c) -1.807 eV,  1.9\%; 
(d) -1.810 eV,  0.9\%; 
(e) -1.175 eV,  0.9\%; 
(f) -0.286 eV,  0.7\%; 
(g) -1.173 eV,  0.5\%; 
(h) -1.851 eV,  0.5\%; 
(i) -1.857 eV,  0.5\%; 
(j) -0.309 eV,  0.5\%. 
The channels are described in Table~\ref{tab_iv_cont}.}
  \label{fig_channel_res}
\end{figure}

\clearpage


\small{
\bibliography{references.bib}
\bibliographystyle{references}
}